\documentclass[pdftex,twocolumn,apj]{aastex63}
\usepackage{color}
\usepackage{amsmath}

\newcommand{\pc}{{\rm pc}}
\newcommand{\kpc}{{\rm kpc}}
\newcommand{\cm}{{\rm cm}}

\newcommand{\Myr}{{\rm Myr}}
\newcommand{\kms}{{\rm km\,s^{-1}}}
\newcommand{\FeH}{{\rm[Fe/H]}}
\newcommand{\OFe}{{\rm[O/Fe]}}
\newcommand{\Msun}{{\rm M}_\odot}

\begin{document}
\shorttitle{Dynamical Evolution of Analogue Magellanic System}

\shortauthors{Williamson \& Martel}

\title{The Evolution of Magellanic-like Galaxy Pairs and The Production of Magellanic Stream Analogues in Simulations with Tides, Ram Pressure, and Stellar Feedback}

\author{David Williamson}
\affil{Department of Physics \& Astronomy, University of Southampton, Southampton, SO17 1BJ, UK}
\email{d.j.williamson@soton.ac.uk}

\author{Hugo Martel}
\affil{D\'epartement de physique, de g\'enie physique et d'optique,
Universit\'e Laval, Qu\'ebec, QC, G1V 0A6, Canada\\
and Centre de Recherche en Astrophysique du Qu\'ebec, C. P. 6128,
Succ. Centre-Ville, Montr\'eal, QC H3C 3J7, Canada}
\email{hmartel@phy.ulaval.ca}

\begin{abstract}
We present a series of chemodynamical simulations of Magellanic-like systems consisting of two interacting, equal-mass dwarf galaxies orbiting a massive host galaxy, including feedback and star formation, tides, and ram pressure. We study the star formation and chemical enrichment history of the dwarfs, and the production of a Magellanic Stream analogue. The dwarfs interact with each other through tidal forces, distorting their morphologies and triggering star formation. A stream is naturally produced as outflows, induced by feedback and interactions, are stretched by tidal forces. Counter to some recent simulations, we find that the leading arm persists even in the presence of ram pressure from the host galaxy. Interactions between the dwarfs and the host galaxies produce multiple kinematic components in the stream, as observed. A combination of ongoing star-formation and entrained low-metallicity gas causes the stream to have a complex chemical structure, with an average metallicity that is generally lower than that of the dwarfs.

\end{abstract}



\section{Introduction}\label{section_intro}

The Magellanic Clouds (MCs) --  consisting of the Large Magellanic Cloud (LMC) and the Small Magellanic Cloud (SMC) -- are the two most prominent and most studied satellite galaxies of the Milky Way (MW). 
They have complex, irregular morphologies, both in the stellar and neutral gas components, which is interpreted as a sign of interaction.
The MCs are gas-rich and metal-poor compared to the MW. The gas fractions (ratio of gas mass to total baryonic mass) are 13\% and 57\% for the LMC and SMC, respectively \citep[][see their Table 1]{df16}. The metallicities of the LMC and SMC are lower than the MW interstellar metallicity by $0.2\,\rm dex$ and $0.6\,\rm dex$, respectively \citep{rd92,rtd02}.

The extended gaseous structure associated with the MCs, the Magellanic Stream (MS), is spectacular evidence of the role of interaction in the evolution of the MCs. The MS 
stretches over $200$ degrees around the MW \citep{1972ApJ...173L.119W,2003ApJ...586..170P,2010ApJ...723.1618N}. Its morphology suggests that the MS is composed of gas that was extracted from the MCs. Furthermore, measurements of the metal abundances are consistent with the MS originating from either the LMC or SMC \citep{2000AJ....120.1830G}. The MS can be divided into a leading arm (LA) and a trailing arm (TA). The LA has a large opening angle, and contains several substructures \citep{2012A&A...547A..12V}. Recent measurements of chemical abundances suggest that the LA is composed of gas that originated in the SMC, and appear to rule out the possibility of an LMC origin \citep{2018ApJ...854..142F}. 

The MCs are also connected by a mostly gaseous bridge, with a stellar fraction of order 0.01\% \citep{df16}. The existence of this bridge is further evidence of interaction between the clouds. The stellar kinematics of the bridge reveal that it is stretching, suggesting that it is tidal forces between the MCs that drive the formation of the bridge \citep{2020arXiv200603163S}. For a recent review of the properties of the Magellanic system, we refer the reader to \citet{df16}.

\subsection{Tidal origin of Magellanic Stream}

Though observations and measurements suggest that the MS is composed of gas extracted from the MCs, the nature of the actual process is not clear. Ram-pressure stripping caused by the MCs' motion through the hot gaseous halo of the MW could explain the origin of the TA, but could not explain the origin of the LA, which, as its name indicates, is a leading structure. Tidal interaction with the MW (and between the MCs) can explain the origin of the MS \citep{2006MNRAS.371..108C,2012ApJ...750...36D,2012A&A...547A..12V,2014MNRAS.443..522Y,2018ApJ...857..101P}. These tidal models could also explain the irregular morphologies of the MCs, and the presence of a bridge between them \citep{ruzickaetal10,guglielmoetal14,mackeyetal18}. They can also produce both the LA and TA with a sufficiently large density and extent, and with the correct orientation. All models that successfully reproduce these structures require at least one, and possibly two passages of the MCs near the MW. \citet{df16} point out that tidal models based on a single passage fail to reproduce the filamentary structure of the MS.

\subsection{Ram-pressure origin of Magellanic Stream}

While models that only include tidal interaction can produce a trailing and a leading structure under the right conditions, they may fail to reproduce several observed properties. Measurements of velocity gradients along the stream indicate that gas extracted from the MCs slows down as it moves away from them, the opposite of what might be expected from tides alone \citep{2012A&A...547A..12V}. 
Also, the chemical abundances of the LA match those of the SMC,
while the kinematics suggest that the LA originated from the LMC, something difficult to explain from tides alone \citep{2018ApJ...854..142F,2019MNRAS.488..918T}.

Consequently, attention has been shifted to models that include a gaseous halo, and the effects of ram pressure \citep{2015ApJ...813..110H,2015ApJ...813...94T,2019MNRAS.488..918T,2019MNRAS.486.5907W}. Ram pressure could either reshape the MS once the gas has been extracted from the MCs by tidal forces, as suggested by \citet{2012A&A...547A..12V}, or it could contribute to the actual extraction of the gas. These models are significantly more complex than purely tidal models, because they depend heavily on the detailed structure of the MW halo, and not merely on the MW mass. The strength of ram pressure, and its effectiveness in stripping gas from dwarf galaxies, depends on the orbital velocities of the dwarfs and the density of the gaseous halo inside which they are moving.

The orbital velocities depend on the total mass of the host galaxy, that can be determined fairly accurately through dynamical measurements. The density, or total mass, of the gaseous halo can be estimated by measuring absorption in spectra of distant QSOs, and the results have been a matter of debate. \citet{2017ApJ...837..169P} conducted a survey of 13 nearby $\rm L_*$ galaxies (the ``COS-Halos'' sample), and found cumulative halo gas masses of order $\sim8\times10^{10}\Msun$ at a radius of $100\,\kpc$, and $\sim10^{10}\Msun$ within $50\,\kpc$. This is considerably higher than an earlier analysis of 
lines of sight through the MW alone \citep{2013ApJ...770..118M}, which found a halo gas mass of $3.8\times10^{10}\Msun$ at a radius of $200\,\kpc$. A subsequent analysis \citep{2018ApJ...862....3B} of the COS-Halos sample, the ``Stock-Bowen'' samples, and additional MW  
lines of sight agrees with the lower gas mass for $\rm L_*$ and MW halos, although with a flatter profile than that found by \citet{2013ApJ...770..118M}.

\subsection{Ejection of gas to form Magellanic Stream}

While tides, ram-pressure, or a combination of both can possibly strip gas located in the outer regions of the MCs, these processes might not be powerful enough to extract gas located deep inside the potential wells of the MCs. \citet{2015ApJ...815...77S} found, in their simulations, that ram-pressure stripping could account for only a few percent of the mass of the MS. By using a denser MW halo, the simulations of \citet{2015ApJ...813...94T} extracted more gas from the MCs, but this gas then dissolves into the halo. Also, if ram pressure is sufficiently strong to form the TA, it might affect the morphology of the LA or even prevent its formation. Essentially, in all such models, ram pressure must be strong enough to extract the gas from the MCs, yet not so strong as to destroy the MS once the gas has been extracted, and these two objectives seem difficult to reconcile. 

If ram pressure is kept at a sufficiently low level to allow the formation of the LA and to prevent the evaporation of the MS into the halo, it might then need some ``assistance,'' to help extracting the gas from the MCs in the first place. This could result from internal processes within the MCs, or from their mutual interaction. \citet{2010ApJ...721L..97B,2012MNRAS.421.2109B} have shown that the influence of the MCs on each other may be more effective than the MW's influence. Here the one-armed spiral and warped stellar bar of the LMC are produced by a close encounter with the SMC, and the MS is produced by the LMC tidally stripping the SMC. \citet{2012ApJ...750...36D} similarly found that recent close encounters between the MCs could play a critical role by ejecting the gas that forms the MS.

Also, several authors have suggested that galactic outflows resulting from stellar formation and feedback could push the gas out of the potential wells. This can produce a ``Magellanic Corona'' \citep{2020hst..prop16363D} which can then be stripped by tides or ram pressure \citep{2018ApJ...854..142F,2018ApJ...857..101P}. Simulations have shown that supernovae-driven outflows can be quite efficient in removing gas from dwarf galaxies \citep{1999MNRAS.309..161M}, and this process is the likely explanation for the stellar-mass/halo-mass relation at low masses \citep[e.g.][]{2015ApJ...799..130R}. With the inclusion of galactic outflows in the models, the ram pressure need not to be as high, and this might help reproducing the morphology and kinematics of the LA, and prevent the dissolution of the MS into the halo.
\subsection{Objectives}

In this present study, we are not attempting to reproduce the Magellanic system specifically. Instead, we focus on the broader issue of the evolution of {\it Magellanic-like} systems composed of two interacting dwarf galaxies orbiting in the vicinity of a massive host galaxy. Dwarf companions are commonly found in numerical simulations of hierarchical structure formation in $\Lambda$CDM Universe \citep{2013MNRAS.428..573S,2018MNRAS.480.3376B,2018ApJ...867...19K}. Though few double dwarfs systems are known at present, some have indeed been observed. \citet{2018MNRAS.480.3376B} studied the distribution of low redshift ($z<0.0252$) dwarfs in SDSS and found that roughly one out of 25 dwarf has a companion of comparable mass. They limited their study to dwarfs that are isolated from more massive galaxies, but while the galaxy mass distribution may be different in richer environments \citep[e.g.][]{2010ApJ...712..484F}, we should still expect to find dwarf pairs. 

Our objective is to understand the role played by star formation and feedback, galactic outflows, tides, and ram pressure in the evolution of Magellanic-like systems, and how the interplay between these various processes determines the morphology, kinematics, and chemical properties of the resulting structures. In an earlier paper (\citealt{2018ApJ...867...72W}, hereafter WM18), we presented a study of the evolution of a single dwarf galaxy orbiting within the gaseous halo of a massive host galaxy. We found that the combined effects of tides and ram pressure shape the morphology and determine the metallicity of the outflows, but have little effect on the final metallicity of the dwarf galaxy itself. In this paper, we extend this early work to the case of two interacting dwarf galaxies orbiting within the halo of a common massive host galaxy. This is the highest resolution numerical study of the evolution of Magellanic-like systems that includes the combined effects of galactic outflows, tides, and ram pressure. The remainder of this paper is organized as follows: In Section~2, we describe the numerical algorithm and the initial conditions used for the simulations. Results are presented in Section~3 and their implications are discussed in Section~4. Our conclusions are presented in Section~5.

\section{Method}\label{section_method}

\subsection{Numerical algorithm}\label{GCDP}

Our simulations are performed using the simulation code unmodified from WM18, but with the initial conditions altered to include two dwarf galaxies, and with other run-time parameters modified. A fuller description of the algorithms are given in our previous papers \citep[][WM18]{2016ApJ...822...91W,2016ApJ...831....1W}, but here we provide a brief summary of the code, directly quoting from WM18.

We use a version of the GCD+ smoothed-particle hydrodynamics (SPH) code \citep{2003MNRAS.340..908K,2012MNRAS.420.3195B,2013MNRAS.428.1968K,2014MNRAS.438.1208K}. This code includes a stochastic star formation model that relaxes the single stellar population assumption, allowing different star particles to represent stars of different masses. Star particles return energy and metals to the ISM through supernovae and stellar winds. We do not include stellar photoionization and radiation pressure, and may somewhat overpredict the star-formation rates for our simulated galaxies \citep{2014MNRAS.445..581H}.

Chemical reactions (e.g. formation of H$_2$) are not explicitly calculated during the simulation (unlike e.g. \citealt{2014MNRAS.443..522Y}), but cooling rates are pretabulated with CLOUDY using a full chemical network. The cooling rates depend on the metallicity, density, and temperature of the gas.

Smoothing lengths are calculated dynamically through an iterative method, so that each particle has $\approx58$ neighboring particles. We apply a minimum smoothing length of $2\,\pc$, which means that particles in very dense regions have $>58$ neighbors. We set the Plummer-equivalent force softening length to $2\,\pc$. 

The metal content of particles is tracked throughout the simulation, and a sub-grid diffusion model allows metals to spread between particles, tracking $8$ separate species. Our version of GCD+ includes modified algorithms for metal deposition and diffusion, as described in \citet{2016ApJ...822...91W}, and a dynamic background potential to represent the varying tidal forces on a satellite galaxy moving through a host galaxy potential, as described in \citet{2016ApJ...831....1W}.

The center-of-mass of the dwarf galaxy system is stationary in our simulation frame, and the host potential is applied as a tidal force, with the position of the host halo directly integrated. To avoid the computational expense of modeling the entire host gaseous halo, we 
treat it using a ``moving box'' method, where halo gas particles are generated or deleted on the boundaries of a cubic box centered on the center-of-mass of the dwarf galaxy system. This allows halo gas particles to have the same mass resolution as the dwarf galaxy particles, preventing a source of numerical error. The temperature and density of 
 particles entering the box through the leading edge are calculated from an analytic profile based on the location of the dwarf system. Gas particles 
 located on the side edges are frozen to maintain (rough) hydrostatic equilibrium in the host halo. This method is described in further detail in WM18, and a similar method was introduced in \citet{2015A&A...582A..23N}. Here we have increased the box width from $160$ kpc to $320$ kpc to capture both dwarfs and the gas stream. 

\subsection{Simulations}\label{section_simulations}

\subsubsection{Host Model}\label{hostmodel}

We use the same MW-like host galaxy model and parameters as Run B of WM18. The host galaxy consists of a dark matter halo of mass $10^{12}\Msun$ with an NFW profile \citep{1997ApJ...490..493N} with a concentration parameter $c=12$. It also contains a gaseous component with a $\beta$-profile, as used in \citet{2015ApJ...815...77S}, with the temperature profile given in \citet{1998ApJ...497..555M}. In our simulations, the dwarf galaxies are never close to the core of the host galaxy, and so we approximate the $\beta$-profile using the large $r$ limit as
\begin{equation}
n(r)=n_{0}\left(\frac{r}{r_c}\right)^{-3\beta}
\label{beta_profile}
\end{equation}
The parameters for the MW $\beta$-profile are taken from \citet{2013ApJ...770..118M}, which is on the lower end of Milky Way gas halo densities. In WM18 we performed simulations with both a MW gas halo, and a gas halo with densities scaled down by a factor of $10$. In both models, both a leading arm and a trailing arm were produced from a single Magellanic-like dwarf. To reduce computational expense while maintaining consistent baryon particle masses between both the dwarfs and the MW halo, most of the simulations in this work use the downscaled halo model, setting $n_0=0.046\,\cm^{-3}$, $r_c=0.35\,\kpc$, and $\beta=0.71$. We do however run a single test (Run~G) at the higher mass, with $n_0=0.46\,\cm^{-3}$, but a shorter simulation time. We do not change the particle mass for Run~G -- the halo is modeled with more gas particles than in runs A--F, and is therefore more computationally expensive. We mostly confine our analysis to the downscaled simulations, Runs A--F, and use Run G as a convergence check. In all these runs, the initial halo metallicity is $10^{-2}\,Z_\odot$. We also perform simulations (Runs H--K) without a host gas halo, to further constrain the effects of ram pressure. Two simulations (Runs J \& K) are also performed without the host gravity field, as a basis to demonstrate the effects of the host galaxy tides.

\subsubsection{Dwarf Galaxy Models}

Our basic dwarf galaxy model is that used in our previous work (\citealt{2016ApJ...822...91W,2016ApJ...831....1W},WM18). This is a Magellanic Cloud analogue, with properties intermediate between the LMC and SMC. Using the same model for both dwarfs allows more direct comparisons with our previous studies, where we modeled a single dwarf galaxy in a tidal field \citep{2016ApJ...831....1W}, and a dwarf galaxy in a tidal field with 
ram pressure (WM18). Unlike other recent work \citep{2012MNRAS.421.2109B,2018ApJ...863...49B,2019MNRAS.488..918T,2019MNRAS.486.5907W}, we are not attempting to model the actual Magellanic system in detail and overfit its present-day configuration (the classic ``weather'' versus ``climate'' problem), but rather to test whether Magellanic-like features can be reproduced even under idealized conditions. Our two dwarf galaxies have identical masses for simplicity. The mass is lower than that of the LMC, as this allows higher resolution, which is critical for resolving feedback \citep[e.g.][]{2006MNRAS.373.1074S,2018MNRAS.477.1578H}.

The initial disk mass for each dwarf is $5\times10^8\Msun$, with a gas fraction of $f_g=0.5$. The stellar disk has a scale height of $100\,\pc$ and a scale length of $540\,\pc$. The gaseous disk has a scale length of $860\,\pc$, and the vertical distribution of gas is initially set by the criterion of hydrodynamic equilibrium, although stellar feedback and radiative cooling cause the gas to rapidly move away from its initial equilibrium state. The initial metal abundances are $[\alpha/\mathrm{H}]=-2$ for all $\alpha$ species, and $\FeH=-3$, giving $[\alpha/\mathrm{Fe}]=1$. The metallicity gradient is initially flat, and so any metallicity gradient produced in the simulations is a result of explicitly-modeled evolution.

Each disk consists of $5\times10^5$ particles, giving a mass resolution of $1000\,\Msun$. This is placed inside an active dark matter halo of mass $9.5\times10^9\Msun$ with an NFW profile with concentration parameter $c=10$, consisting of $9.5\times10^5$ particles. 

\subsubsection{Orbits}

\begin{table*}
\setlength\tabcolsep{4pt} 
    \begin{center}~~~~~~~~~~~~~~~~~~~~~~~~~~~~~~
        \begin{tabular}{cccccccccccc}
        \hline\hline
        Run & $x_{{\rm COM},i}^{\phantom i}$ & $y_{{\rm COM},i}^{\phantom i}$ & $v_{x,{\rm COM},i}^{\phantom i}$ & $v_{y,{\rm COM},i}^{\phantom i}$ & $\Delta x_i$ & $\Delta y_i$ & $\Delta v_{x,i}$ & $\Delta v_{y,i}$ & $r_{{\rm per},1}$ & $r_{{\rm per},2}$ & $n_0$\\
        ~ & [kpc] & [kpc] & [km/s] & [km/s] & [kpc] & [kpc] & [km/s] & [km/s] & [kpc] & [kpc] & [cm$^{-3}$]\\
        \hline
    A & $0$ & $-220$ & $-60$ & $120$ & $0$ & $20$ & $50$ & $0$ & $61$ & $49$ & $0.046$\\
    B & $0$ & $-230$ & $-85$ & $120$ & $0$ & $20$ & $50$ & $0$ & $89$ & $82$ & $0.046$\\
    C & $0$ & $-110$ & $-190$ & $0$ & $0$ & $20$ & $-30$ & $0$ & $96$ & $96$ & $0.046$\\
    D & $0$ & $-110$ & $-190$ & $0$ & $0$ & $20$ & $-50$ & $0$ & $87$ & $79$ & $0.046$\\
    E & $-100$ & $-10$ & $0$ & $190$ & $0$ & $20$ & $30$ & $0$ & $100$ & $97$ & $0.046$\\
    F & $-100$ & $-10$ & $0$ & $190$ & $0$ & $20$ & $50$ & $0$ & $102$ & $64$ & $0.046$\\
    G & $0$ & $-110$ & $-190$ & $0$ & $0$ & $20$ & $-50$ & $0$ & $86$ & $79$ & $0.460$\\
    H & $0$ & $-230$ & $-85$ & $120$ & $0$ & $20$ & $50$ & $0$ & $89$ & $82$ & $0$\\
    I & $0$ & $-110$ & $-190$ & $0$ & $0$ & $20$ & $-50$ & $0$ & $87$ & $79$ & $0$\\
    J & $-$ & $-$ & $-$ & $-$ & $0$ & $20$ & $50$ & $0$ & $-$ & $-$ & $0$\\
    K & $-$ & $-$ & $-$ & $-$ & $0$ & $20$ & $-50$ & $0$ & $-$ & $-$ & $0$\\

        \hline
        \end{tabular}
    \end{center}
\label{init_table}
\caption{Parameters of the simulations. Note that $z=v_z=0$ for all galaxies initially. 
Subscripts~$i$ indicate initial values.
Pericenters are directly measured from the simulation. Runs J and K do not show COM distances from host halo center as these are performed in isolation
}
\end{table*}

\begin{figure}
\begin{center}
\includegraphics[width=\columnwidth]{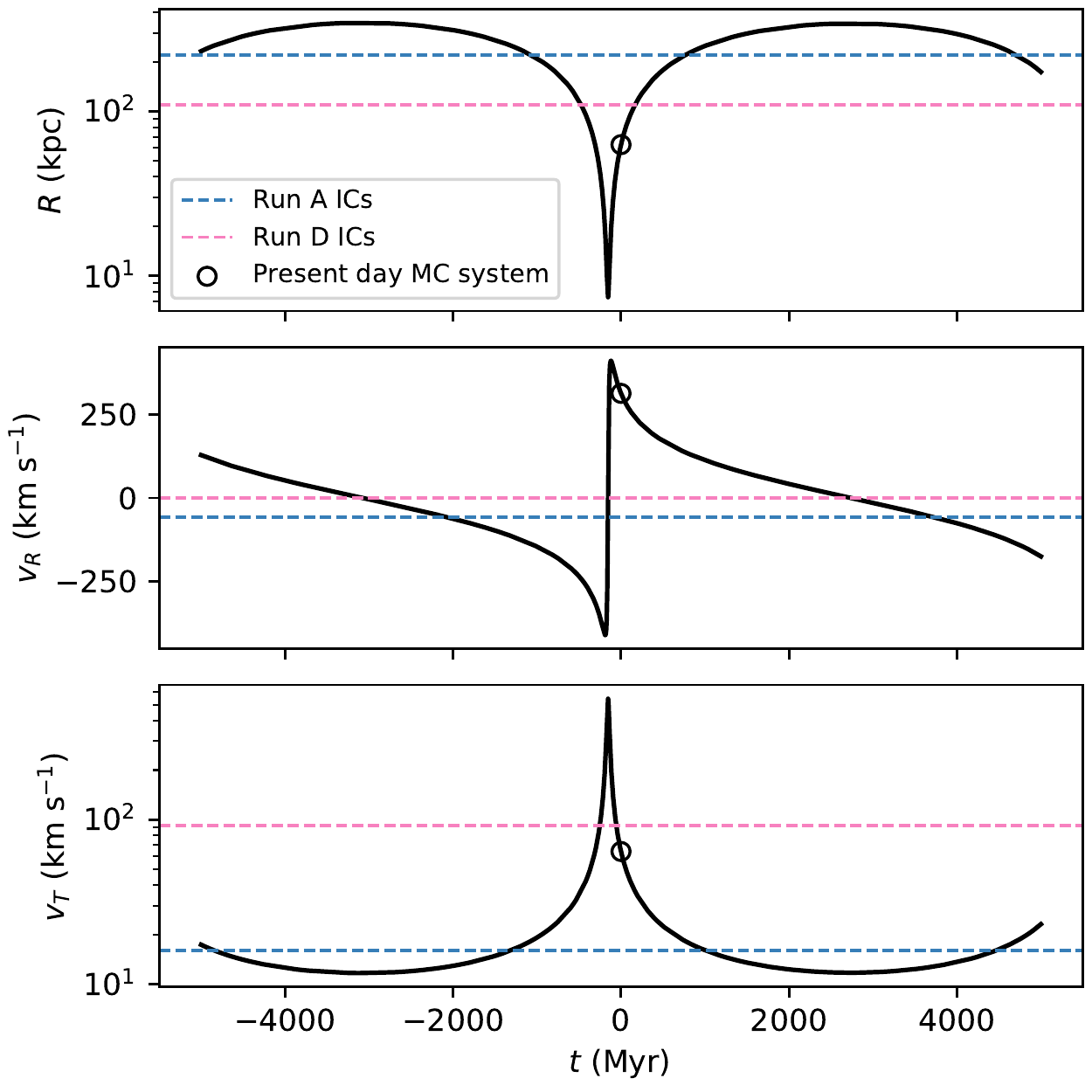}
\end{center}
\caption{\label{orbit_ICs} Integration of MC positions and velocities, showing present day positions of MCs, and initial conditions for Run A (infall) and Run D (circular orbit). Top: distance from MW center. Middle: Radial velocity. Bottom: Tangential velocity.}
\end{figure}

\begin{figure*}
\begin{center}
\includegraphics[width=\textwidth]{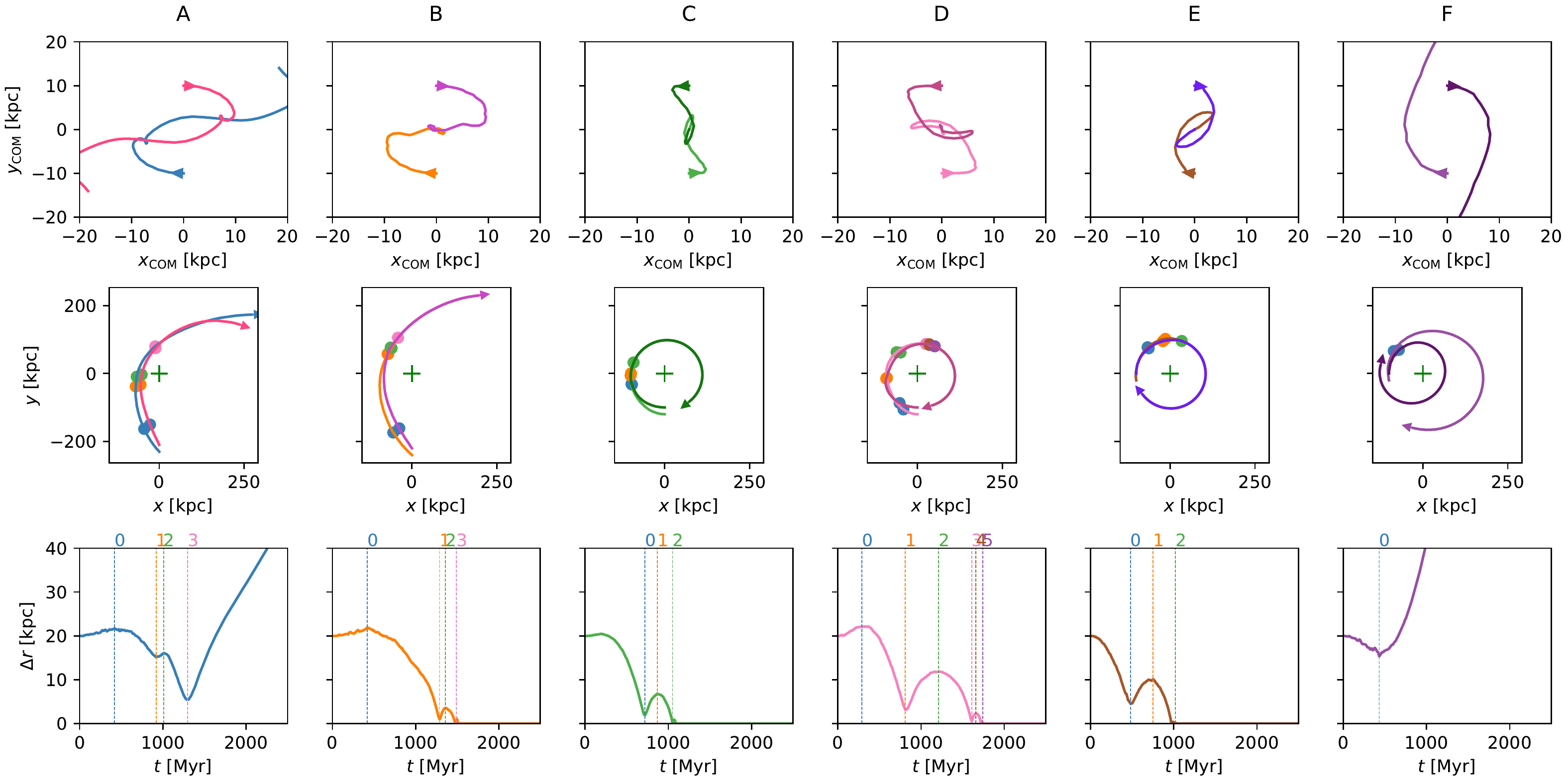}
\end{center}
\caption{\label{timesummary} Orbital characteristics for Runs A-F. Top row: orbits of dwarfs in dwarf center-of-mass frame. Arrows indicate direction of initial velocities. Middle row: orbits of dwarfs in host frame. Local maxima and minima of dwarf-dwarf orbital distances ($\Delta r$) are indicated with colored points, corresponding to dashed vertical lines in the time-series data in the bottom row (and in Figure~\ref{timesummary2}). Crosses indicate the location of the central galaxy. Bottom row: dwarf-dwarf orbital distance versus time.}
\end{figure*}

\begin{figure*}
\begin{center}
\includegraphics[width=0.95\textwidth]{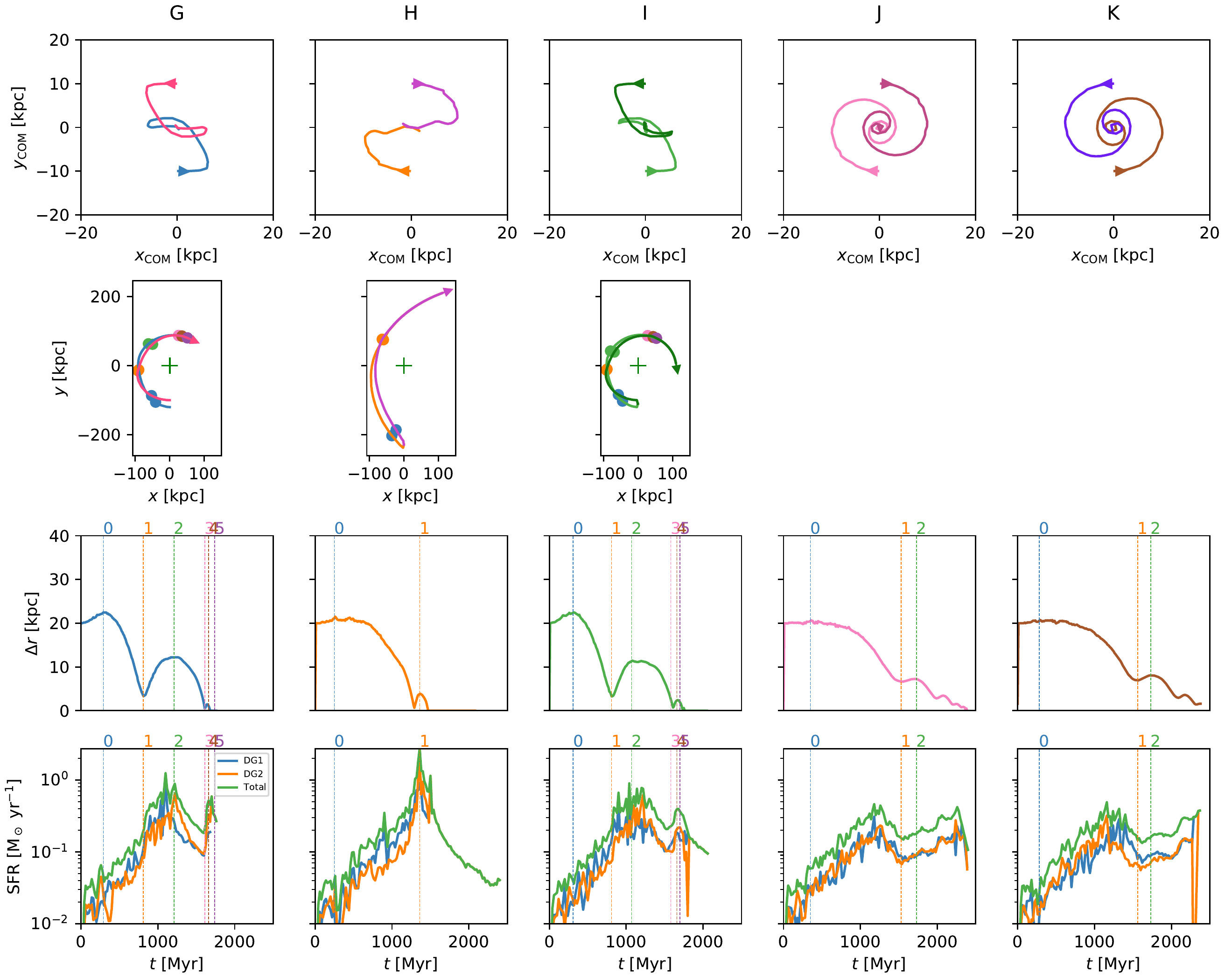}
\end{center}
\caption{\label{extratimesummary} Orbital characteristics and star formation rates for Runs G-K. Top row: orbits of dwarfs in dwarf center-of-mass frame. Arrows indicate direction of initial velocities. Second row: orbits of dwarfs in host frame. Local maxima and minima of dwarf-dwarf orbital distances ($\Delta r$) are indicated with colored points, corresponding to dashed vertical lines in the time-series data in the third row. Crosses indicate the location of the central galaxy. Third row: dwarf-dwarf orbital distance versus time. Fourth row: star formation rates for dwarfs}
\end{figure*}

Proper motion studies suggest that the MCs have had a close encounter with each others in the past \citep{2018ApJ...864...55Z}, and may be approaching the MW for the first time \citep{2006ApJ...652.1213K,2006ApJ...638..772K,2007ApJ...668..949B,2008AJ....135.1024P}. To determine MC-like initial conditions for the DGs orbit around the host halo, we take the LMC and SMC position and velocities of \citet{2018ApJ...864...55Z} and \citet{2013ApJ...764..161K}, and integrate this motion forwards and backwards by $5000$ Myr, as shown in Figure~\ref{orbit_ICs}. The exact details of this motion will depend on the Milky Way potential and MC proper and radial velocities, but here we are interested in indicative initial conditions and not a precise model of the exact history of the MCs. We use these as the basis for two sets of dwarf orbits around the host.

We base one set of ICs on the predicted position and velocity of the MCs about $1500-2000$ Myr in the past, where the dwarfs are initially $\sim220\,\kpc$ from the MW center. We slightly tweak these initial conditions from our predicted past coordinates, to increase the pericenter of the DGs orbit. Our predictions showed a pericenter of $<10$ kpc, which would heavily disrupt the system and not produce MC-like galaxies. This setup is used for Runs A, B, \& H. We plot the initial coodinates for Run A in Figure~\ref{orbit_ICs} for comparison with our predicted MC past orbits.

The second set of ICs is a circular orbit with a radius of $\sim110\,\kpc$ which allows us to investigate the effects of tides and ram pressure over an extended period of time, in an idealized environment. In all simulations, the host galaxy is at rest at the origin. This setup is used for Runs C--G \& I.  We plot the initial coodinates for Run D in Figure~\ref{orbit_ICs} for comparison with our predicted MC past orbits.

As our DGs do not have a mass ratio similar to that of the MCs, we are not able to directly take our dwarf-dwarf orbital properties from the MCs. Instead, we produced a series of low-resolution test runs to explore which dwarf-dwarf orbits produced MC-like interactions. For simplicity, the orbits of the dwarfs around each other are co-planar with their orbit around the host galaxy, which we define as the $x$-$y$ plane, i.e. initially $v_z=0$ and $z=0$ for all galaxies. Both dwarfs rotate in the same direction, inclined $90^\circ$ from the orbital planes. We vary the dwarf-dwarf relative tangential velocity and relative position in these test runs, while keeping their initial separation constant at $\Delta r=20\,\kpc$ and initial relative radial velocities at zero. We discarded runs where the dwarfs immediately merged, or quickly escaped from each other without interaction. It appears that the specific present-day configuration of the Magellanic Clouds, as a pair of strongly interacting dwarf galaxies, under strong influence of a host galaxy, may be a rare situation. As the strong interactions appear to be short-lived before the dwarfs escape each other or merge, some fine-tuning is required to ensure that the timing is correct. From our test runs, we found the resulting ``interesting'' simulations had dwarf-dwarf relative velocities of $\pm30\,\kms$ and $\pm50\,\kms$, where a positive or negative velocity indicates that the dwarf's orbit (the orbit  of the dwarfs around each other) is prograde or retrograde, respectively, relative to the dwarfs' orbit around the host galaxy. We selected these initial conditions to re-run the simulations at full resolution.

The initial coordinates of both galaxies in all full-resolution runs are summarized in Table~1, as well as the pericenter radius reached by each dwarf galaxy in each simulation. The first column gives the name of each run. Columns 2--5 give the initial position and velocity of the center-of-mass of the dwarf system. Columns 6--9 give the position and velocities of the dwarfs relative to each other. Columns 10 and 11 give the pericenters for each dwarf. The last column gives the gas density $n_0$ used for the density profile of the host galaxy gas halo (eq.~\ref{beta_profile}). The pericenters are directly obtained from the simulations and therefore include any hydrodynamic effects on the galaxy orbits.

To summarize, Runs A, B, \& H represent eccentric orbits around the host galaxy, with the same relative velocity between the dwarfs. In all three runs, the dwarf orbits are prograde with respect to their orbit around the host galaxy, and have the same speed - we found these eccentric orbits were particularly sensitive to parameter variations that would cause the dwarfs to either merge or escape each other almost immediately. Runs C--F, \& I represent five runs with circular orbits around the host, with the dwarf's orbit being either prograde or retrograde, and with two different relative velocities between the dwarfs. As stated in Section~\ref{hostmodel}, Run G is similar to Run D but with a higher host halo gas density, while the mass resolution of Run G is the same as in the other runs - the run has more particles and is therefore more computationally expensive. Runs H--K are performed without a host gas halo, and Runs J \& K are performed without the host gravity field - i.e. these are simulations of isolated dwarf pairs. Run H is equivalent to Run B without a gas halo, Run I is essentially Run D without a gas halo. Run J has the same dwarf-dwarf configuration as Runs A, B, F, \& H, but in isolation. Similarly, Run K has the same dwarf-dwarf configuration as Runs D, G, \& I.

We define the center of each dwarf galaxy as the local minimum of the gravitational potential, using the {\sc pykdgrav} python library to recalculate the potential. As we note in Section~\ref{chemo_section}, due to disturbances from interactions and strong feedback, this center may not correspond to other centers of interest, such as the metallicity-slope center

\section{Results}\label{section_results}

\begin{figure*}
\begin{center}
\includegraphics[width=\textwidth]{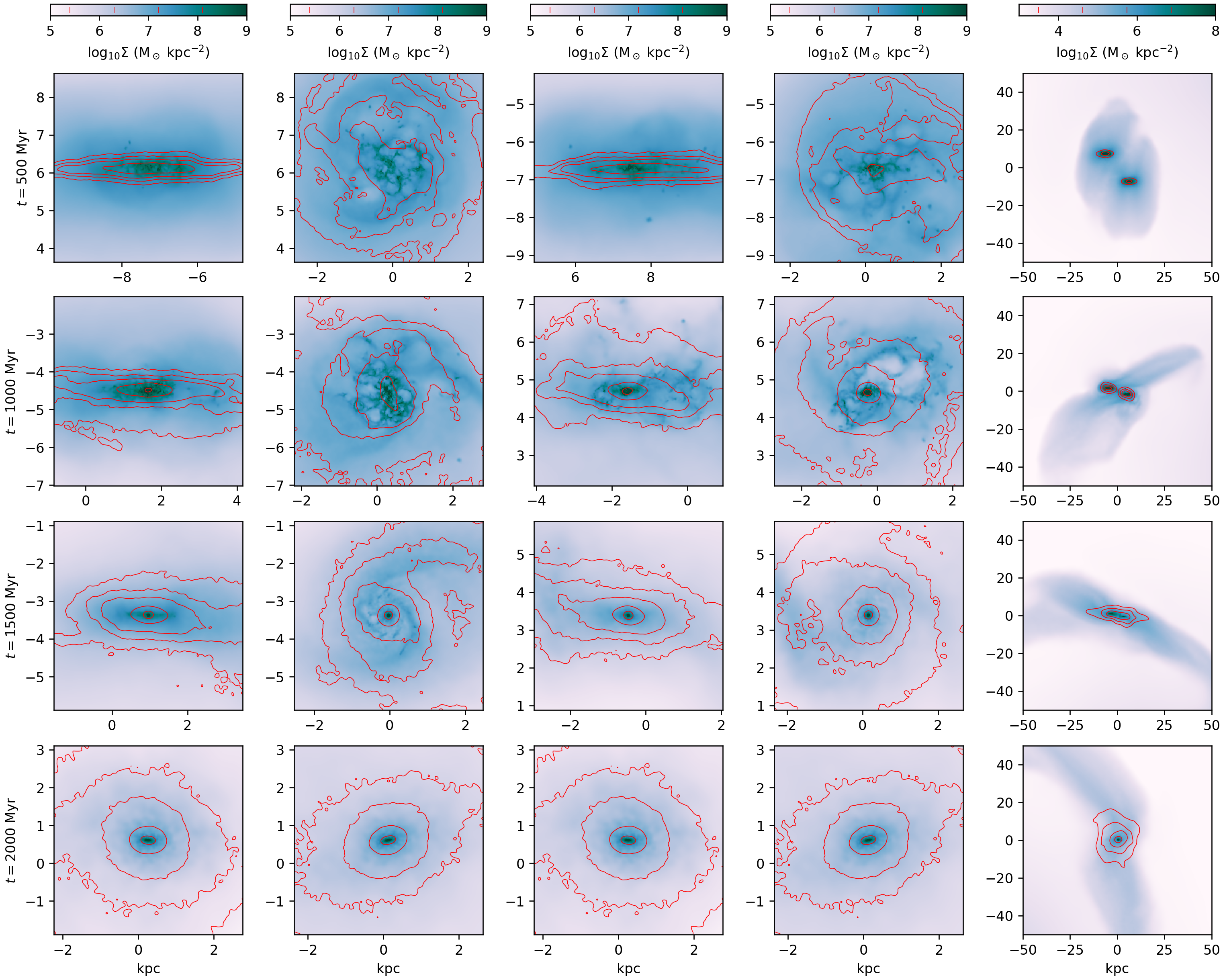}
\end{center}
\caption{\label{D_evolution} Summary of the evolution of the system for Run D. Gas surface density is shown on a blue-gray color-map, stellar surface density is shown with red contours. The values of the contours are shown in the color bars on top). All coordinates are relative to the center-of-mass of the moving box, in kiloparsec units. Columns from left to right: edge-on view of dwarf 1, face-on view of dwarf 1, edge-on view of dwarf 2, face-on view of dwarf 2, edge-on view of entire system. Rows from top to bottom 
correspond to different epochs, as indicated on the left. At $t=2000\,\Myr$, the dwarfs have merged, and the merger remnant is shown twice in the bottom row.}
\end{figure*}

\begin{figure*}
\begin{center}
\includegraphics[width=\textwidth]{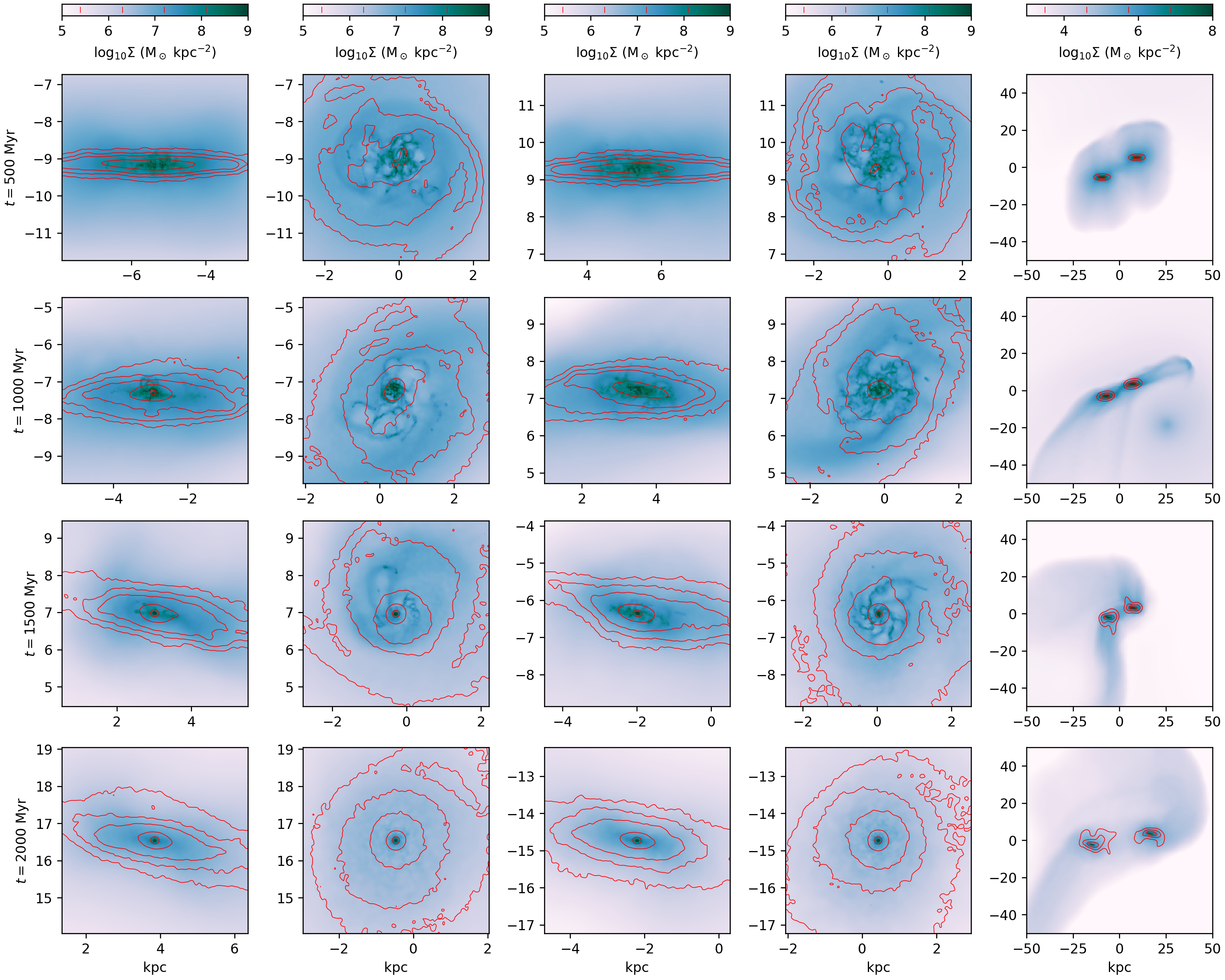}
\end{center}
\caption{\label{A_evolution} 
Same as Figure~\ref{D_evolution}, for Run~A}
\end{figure*}

\begin{figure}
\begin{center}
\vskip11pt
\includegraphics[width=\columnwidth]{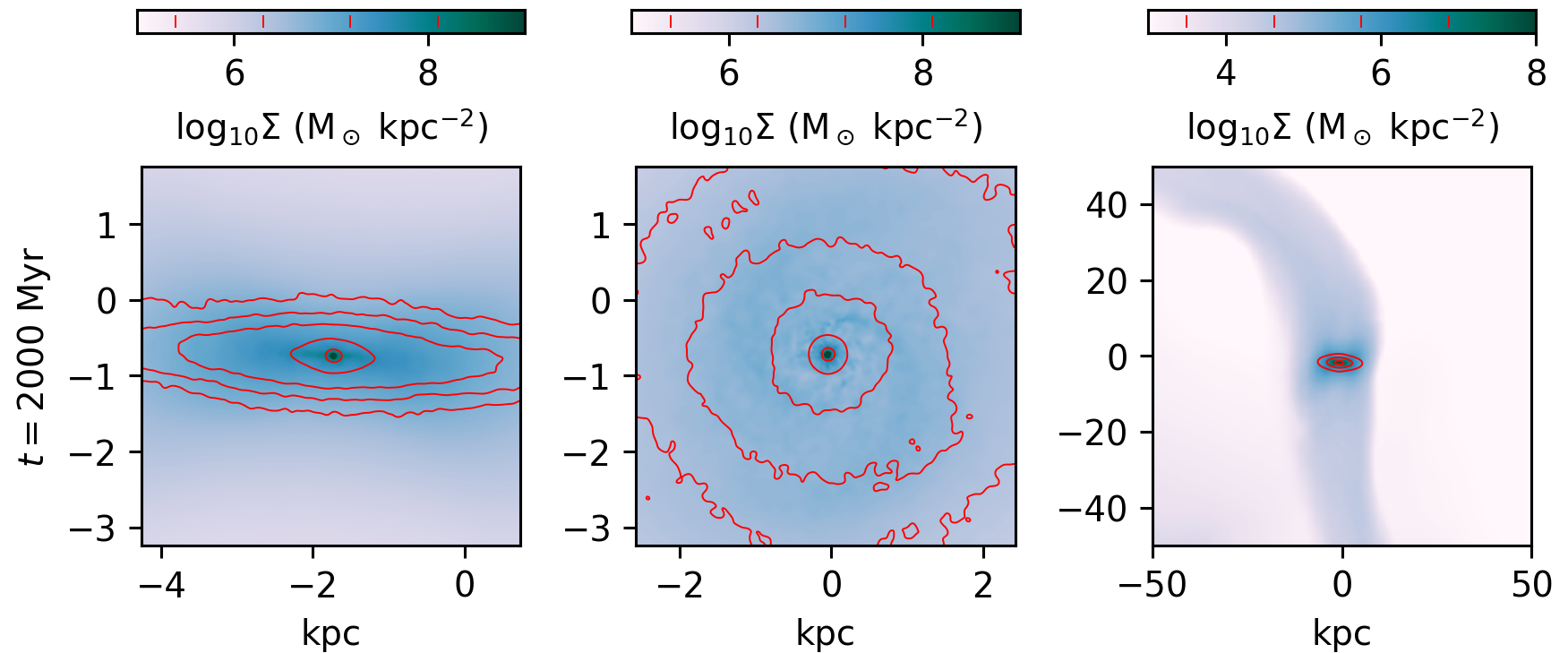}
\end{center}
\caption{\label{oldB_evolution}
Final stage of the evolution of a single dwarf system (Run B of WM18). Colors and contours have the same meaning as in Figures~\ref{D_evolution} and~\ref{A_evolution}.
}
\end{figure}

\begin{figure*}
\begin{center}
\includegraphics[width=\textwidth]{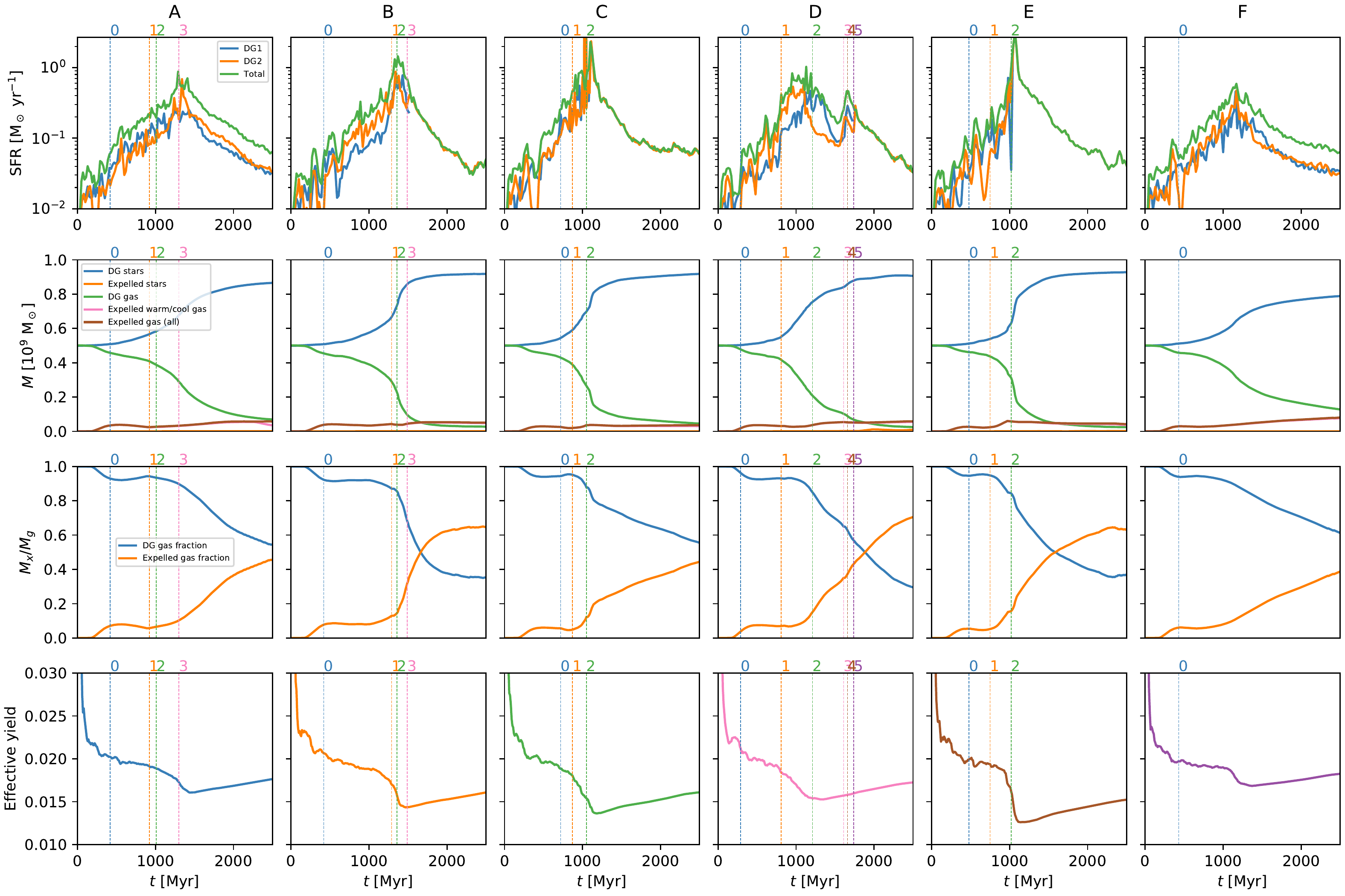}
\end{center}
\caption{\label{timesummary2}
Comprehensive summary of Runs A--F. Top row: star formation rate in each dwarf galaxy (blue and orange lines) and combined star formation rate (green lines), versus time. Second row: relative mass in different phases and environments - blue: stars within dwarfs, orange: stars outside dwarfs, green: gas within dwarfs, pink: ``warm/cool'' gas outside dwarfs, brown: all gas outside dwarfs. Third row: fractions of gas inside (blue) and outside (orange) dwarfs. Fourth row: effective yield (see section~\ref{chemo_section}). Vertical lines indicate ``times of interest'' (see Fig.~\ref{timesummary}).}
\end{figure*}

The characteristics of the orbits for Runs~A--F and Runs~G--K are shown, respectively, in the top three rows of Figures~\ref{timesummary} and \ref{extratimesummary}. The top rows show the trajectories of the dwarfs in dwarf center-of mass coordinates, while the second rows show them in host galaxy center coordinate (notice that Runs~J and~K have no host). The third rows show the separation between the dwarfs over time. On these panels, (and also in all panels of Figure~\ref{timesummary2} below) vertical color lines show ``times of interest,'' where the separation is a prominent local minimum or maximum. We have manually excluded some maxima and minima to avoid overcrowding the plots. The colored dots in the second rows show the locations of the dwarfs at these particular times.

The main simulations run for $2500\,\rm Myr$, which is approximately one circular orbit for Runs C--E. These simulations end more than $1000\,\rm Myr$ after the dwarfs merge or have their nearest encounter. The dwarfs eventually merge in Runs B--E, after they have experienced one or two close encounters. We continue to evolve the simulations after merger, and hence there is only a single trajectory after the merger. Simulations G--K are supplemental simulations used as bases for comparison.

In Runs A and F, the dwarfs escape each other after experiencing at least one close encounter. The dwarfs separate in Run F because their orbit around each other is prograde with respect to their orbit around the host galaxy, with a faster relative velocity than in Run E (where the dwarfs' orbit is also prograde). One dwarf is swung out to a considerably more distant orbit after a single close encounter between the dwarfs. In Run A, the dwarfs pass close to each other twice, but the orientation and timing of their orbits and the host tidal field are such that they just miss each other and then separate. While the dwarfs do not merge in these simulations, they still demonstrate the effects of interactions between dwarf galaxies within a host galaxy halo.

The trajectories of Runs H and I are almost identical to those of Runs B and D, which share initial orbital configurations but include a host gas halo. In Runs J and K, the isolated dwarf pairs merge later than in any other simulation where the dwarfs do not escape each other. In the absence of perturbations from the tidal field, the dwarfs can slowly spiral towards each other, losing orbital energy through dynamical friction. This demonstrates that the orbits of our double dwarf systems (and therefore, the time-scale for any merger) are utterly dominated by gravitational forces - ram pressure and other hydrodynamic effects are not significant.

\subsection{Morphology}

\begin{figure*}
\begin{center}
\includegraphics[width=\textwidth]{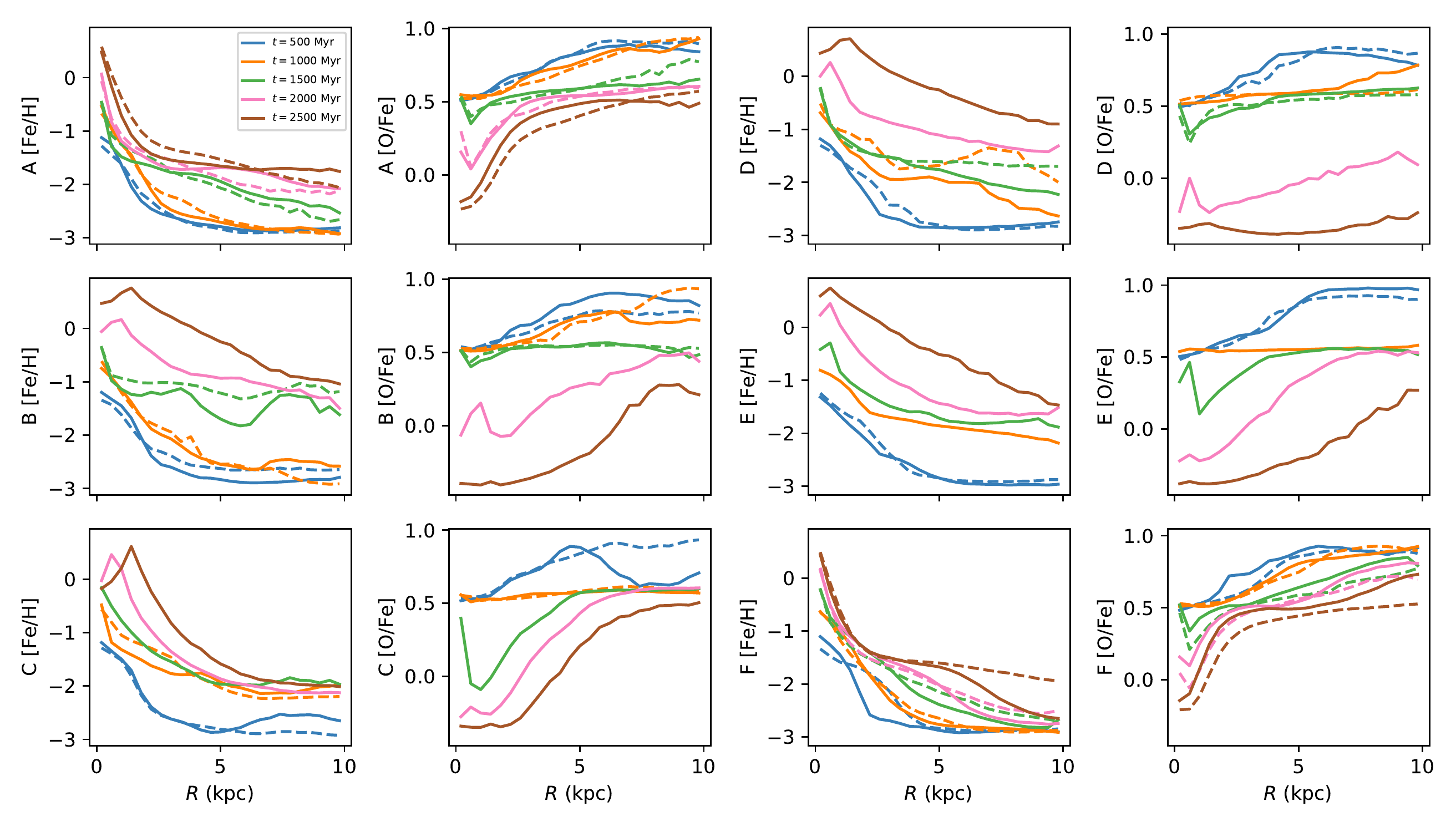}
\end{center}
\caption{\label{profiles}
Radial profiles of abundance ratios in the gas phase of each dwarf galaxy, for Runs~A--C (first and second columns) and Runs~D--F (third and fourth columns), as indicated. $R$ is the radial distance from the minimum of the potential well, in the disk plane. Colors identify the epoch, as indicated. The two different line types identify the two dwarf galaxies. The absence of a dashed line of a particular color indicates that the two dwarf galaxies have merged.}
\end{figure*}

A visualization of the evolution of the system for Run~D is shown in Figure~\ref{D_evolution}.
Results for Runs~B--E are similar. In both dwarfs, star formation proceeds rapidly with strong feedback, producing supernova bubbles at $t=500$ Myr, as seen in the first four panels of the top row. The stellar distribution is irregular at early times as new stars are formed stochastically from the feedback-disturbed gas \citep[cf.][]{2014MNRAS.438.1208K}, though a hint of spiral arm structure may be visible. The feedback expels gas, which forms a halo around each dwarf, as seen in the rightmost panel. Tidal forces between the dwarfs drive some of this gas into a bridge between the dwarfs. At this early stage, much of the characteristic properties of the MC system are already formed.

By $t=1000$ Myr, tidal interaction between the dwarfs becomes more dominant, and tidal arms can be seen in the stellar distribution. The outflows have also been shaped by the tides of the host into a stream. By $t=1500$ Myr, the gas density inside the dwarfs has been reduced, due to expulsion of gas, and consumption by star formation. Fewer supernova bubbles are visible. Tidal stirring also causes the dwarf disks to thicken. Finally, by $t=2000$ Myr, the dwarfs have merged. This remnant is a boxy elliptical galaxy, with long gas streams.

A visualization of the evolution of the system for Run~A is shown in Figure~\ref{A_evolution}. As noted above, Runs A and F do not result in a merger. Here we still see anisotropy in the stellar distribution from stochastic star formation, but tidal tails from the interaction between the dwarfs are less apparent. The dwarfs do still form an extended gas stream, and a gas bridge between them. After their close encounter, both dwarfs settle back into fairly regular disks, but inclined relative to their initial orientation. This inclination is caused by the tidal field of the host galaxy. For further comparison, we show in Figure~\ref{oldB_evolution} the final stage of the evolution, at $t=2000\,\rm Myr$, of a single dwarf system. This is ``Run~B'' from WM18. Again, the gas and stars are disturbed primarily by feedback, which produces an outflow, that is shaped by the tidal field of the host. The dwarf itself settles into a fairly regular disk.

These results suggest that most of the irregularity in the stellar distributions of the dwarfs is caused by the dwarfs themselves -- through feedback and mutual interactions -- and not by the host galaxy. The thickening of the dwarf disks is also caused by interactions between the dwarfs, and eventually by their merger. This is in agreement with the classic picture of the morphological origin of dwarf ellipticals from tidal stirring \citep{1996Natur.379..613M,2001ApJ...559..754M,2010MNRAS.405.1723S}. The host galaxy has a gentler effect on the dwarf morphologies, and only dominates the shape of the outflowing stream.

\subsection{Star formation and outflows}\label{sf_section}

\begin{figure*}
\begin{center}
\includegraphics[width=\textwidth]{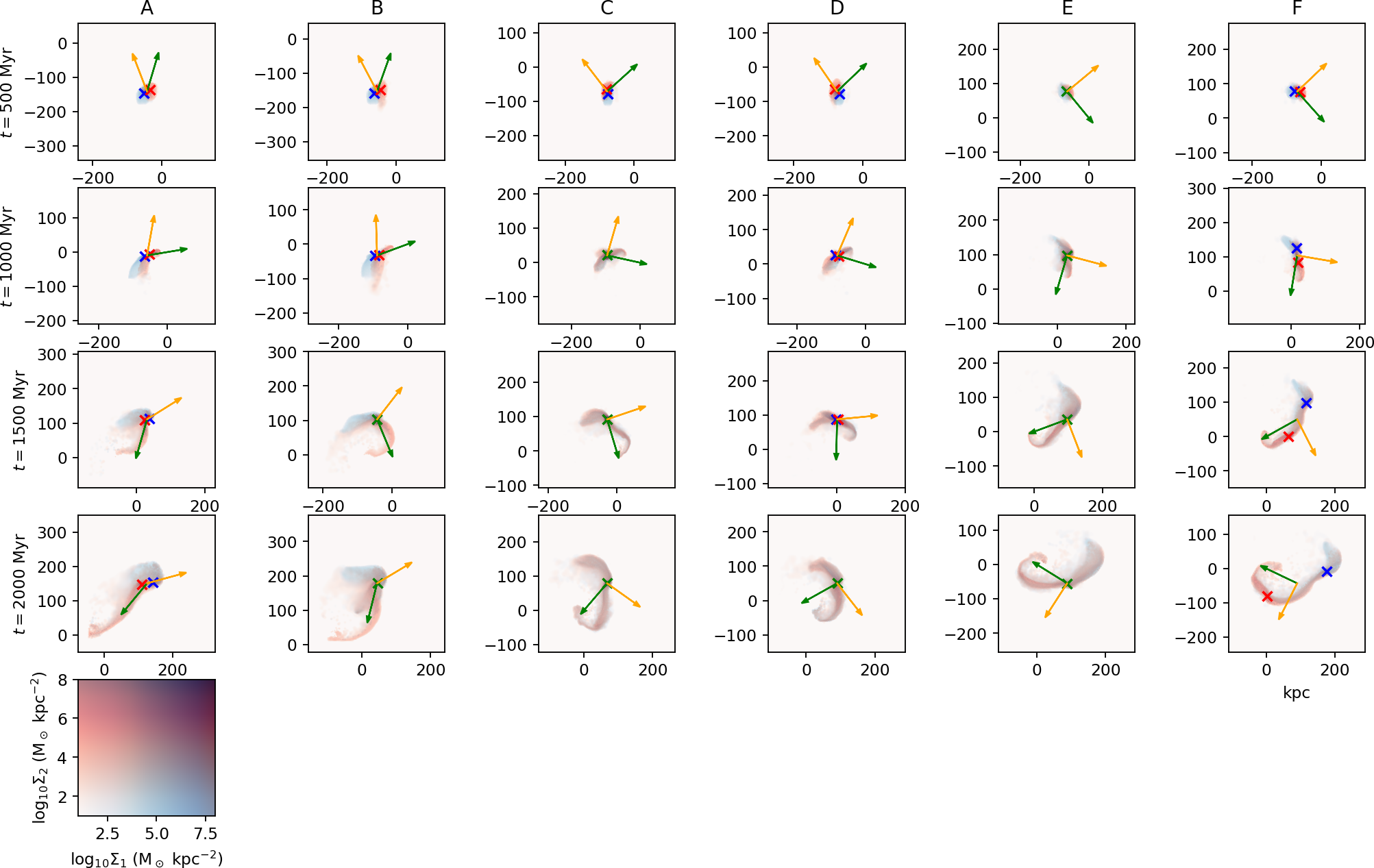}
\end{center}
\caption{\label{column}
Evolution of the gas density, for Runs~A--F. Red and blue 
colors identify, respectively, gas that originated in the first
dwarf and in the second dwarf. Green arrows indicate the direction of the host galaxy center, orange arrows indicate the direction of motion (but not magnitude of the velocity) of the dwarfs around the host, blue and red crosses indicate locations of each dwarf, green cross indicates merged dwarf location. The host halo center is at the origin, and a green arrow indicates the direction of the host center from the dwarfs' center of mass.}
\end{figure*}

\begin{figure*}
\begin{center}
\includegraphics[width=0.66\textwidth]{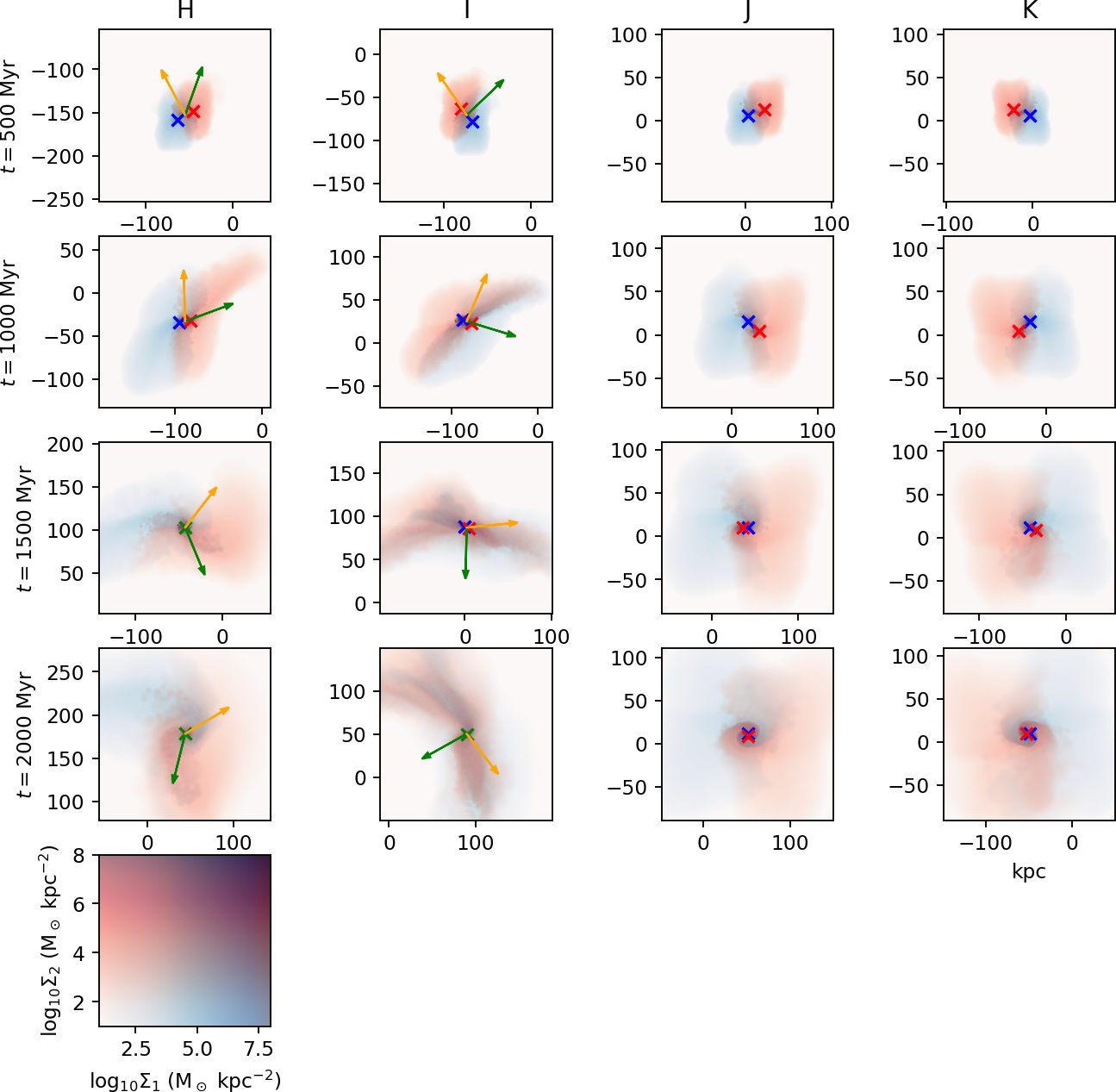}
\end{center}
\caption{\label{column2}
Evolution of the gas density, for Runs~H--K. Red and blue 
colors identify, respectively, gas that originated in the first
dwarf and in the second dwarf. Green arrows indicate the direction of the host galaxy center, orange arrows indicate the direction of motion (but not magnitude of the velocity) of the dwarfs around the host for Runs H \& I, blue and red crosses indicate locations of each dwarf, green cross indicates merged dwarf location. The host halo center is at the origin, and a green arrow indicates the direction of the host center from the dwarfs' center of mass.}
\end{figure*}

The top row of Figure~\ref{timesummary2} and the bottom row of Figure~\ref{extratimesummary} show the star formation rate (SFR) for both dwarfs (blue and orange lines), and the sum of the two (green lines), for Runs A--F and H--K. In Runs~B--E, the dwarfs eventually merge, and from that point only the green line is plotted. The SFR rises to a peak at $t\sim1200\,\rm Myr$, driven by internal gravitational collapse, and by tidal stirring. Star formation then slowly shuts down as the gas is consumed or expelled.

The peak of star formation occurs at roughly the same time in all runs, even in the absence of an external tidal field, suggesting that internal processes dominate star formation -- star formation increases as instabilities form, and then slows down as the fuel has been consumed. The single-dwarf models presented in WM18 all had a similar star-formation peak at $t\sim1200\,\Myr$, as do Runs J \& K in the absence of an external tidal field. Dwarf-dwarf interactions have a significant secondary effect. Runs~B--E, which have close encounters that end in a merger, have a higher peak of star formation than Runs~A and~F. In particular, Run~D shows two star formation peaks, triggered by two close encounters separated by $\sim1000\,\Myr$. In these runs, the final merger results in a starburst, during which a significant fraction of the remaining gas is converted to stars. Ram pressure and host tides seem to have little effect on star formation. The two dwarfs in Run~F move apart as they follow different trajectories, but do not have significantly different star formation rates. Runs~A and~B have similar orbits around the host galaxy, but show significantly different star formation rates due to interactions between the dwarfs. Runs J \& K, in the absence of a tidal field, show similar (low) star formation rates to Run F, as the dwarfs interact gently over a long period of time rather than having strong star-formation triggering encounters. Overall, the picture is that star formation is primarily driven by internal processes, secondarily affected by close encounters with the other dwarf galaxy, and not noticeably affected by the host galaxy, except in how the host galaxy affects the dwarf-dwarf orbits.

We track the masses of stars, and of total and ``warm/cool'' gas for Runs A--F in the second row of Figure~\ref{timesummary2}. Here ``warm/cool'' gas is defined at gas with temperatures below $2\times10^4\rm K$, equivalent to the ``neutral gas'' definition in \citet{2019MNRAS.486.5907W}. We divide the gas and stars into the dwarf galaxy (DG) population, defined as gas and stars within $10\,\pc$ of the center of either dwarf, and ``expelled'' gas and stars 
as gas and stars located outside that radius, but that originated from the dwarfs, not from the host galaxy. 
The mass of ejected stars is quite low, and this (orange) line is not visible in many plots. The (pink) warm/gas line is also not often visible, because almost all of the expelled gas is in this warm/cool phase and the pink and brown lines almost always coincide in the figure -- the vast majority of hot ($\ge2\times10^4\rm K$) gas outside of the dwarfs is that which originated from the host galaxy halo.

The strongest effect on the gas mass of the dwarfs is star formation. The gas in the dwarfs is consumed by star formation and converted into stars. Feedback pushes out only a fraction of the warm/cool gas into the stream, while the rest is consumed by star formation. Further gas is expelled when galaxies have a close encounter or merge, both due to the disruption of the encounter, and due to the resulting burst of star formation. This is more clearly shown in the third row of Figure~\ref{timesummary2}, where we plot the fraction of gas within the dwarfs and expelled from them. Generally, the expelled gas is comparable to or greater than the retained gas by the end of the simulations, consistent with observations of the MS \citep{2014ApJ...787..147F,2017ApJ...851..110B,2017A&A...607A..48R}. Although only a fraction of the initial gas mass is expelled, the consumption of gas by star formation causes the expelled and retained masses to be similar. This suggests that it does not require extremely effective stripping or outflows for the gas mass of the MS to be similar to that of the MCs - modest outflows plus gas depletion by ongoing star formation can achieve this instead.

The final ratio between the expelled gas and retained gas masses can be considered an approximate proxy for the strength of interactions between the dwarfs. In Run~F, the dwarfs only pass by each other early on, the star formation peak is low, and most of the gas is retained in the dwarfs. In Run~A, the galaxies have a longer and closer encounter, and the retained gas fraction is smaller. In Runs~B, D, and~E, the dwarfs undergo at least one close encounter before merging, and most of the remaining gas ends up in the stream. Of course, this is not the only factor -- ongoing star formation will also continue to expel and consume gas. Despite the strong interaction and sharp star formation peak in Run~C, the final expelled gas fraction is lower than that of the similar Runs~D and~E. This is because Runs~D and~E have stronger ongoing star formation in the final 1000 Myr of the simulation.

We can further illustrate the relative importance of the final starburst and instabilities triggered by close encounters by comparing Run~E with Runs B--D. In Run~E, the encounter is not as ``close'' as in Runs B--D, with a minimum separation of $4.5\,\rm kpc$. The resulting effect is weaker, and as a result, the amount of gas remaining before the final merger takes place is larger, resulting in a much stronger starburst. The peak SFR in Run~E is larger than in Run~C, and more than twice as large as in Runs~B and D, resulting in a sharp increase in stellar mass at $t=1000\,\rm Myr$. We note that the competing effects of instabilities and mergers conspire to produce a comparable stellar mass by the end for the simulations.

Both star formation and interactions between the dwarfs are of similar importance for driving outflows, while host tides and ram pressure have only a small effect (beyond the host perturbing the dwarf orbits and modulating their interactions somewhat). Dwarf-dwarf interactions are effectively counted twice here - the interactions can directly expel material through tidal forces, and indirectly by triggering bursts of star formation.

\subsection{Internal Chemodynamics}\label{chemo_section}

\begin{figure*}
\begin{center}
\includegraphics[width=\textwidth]{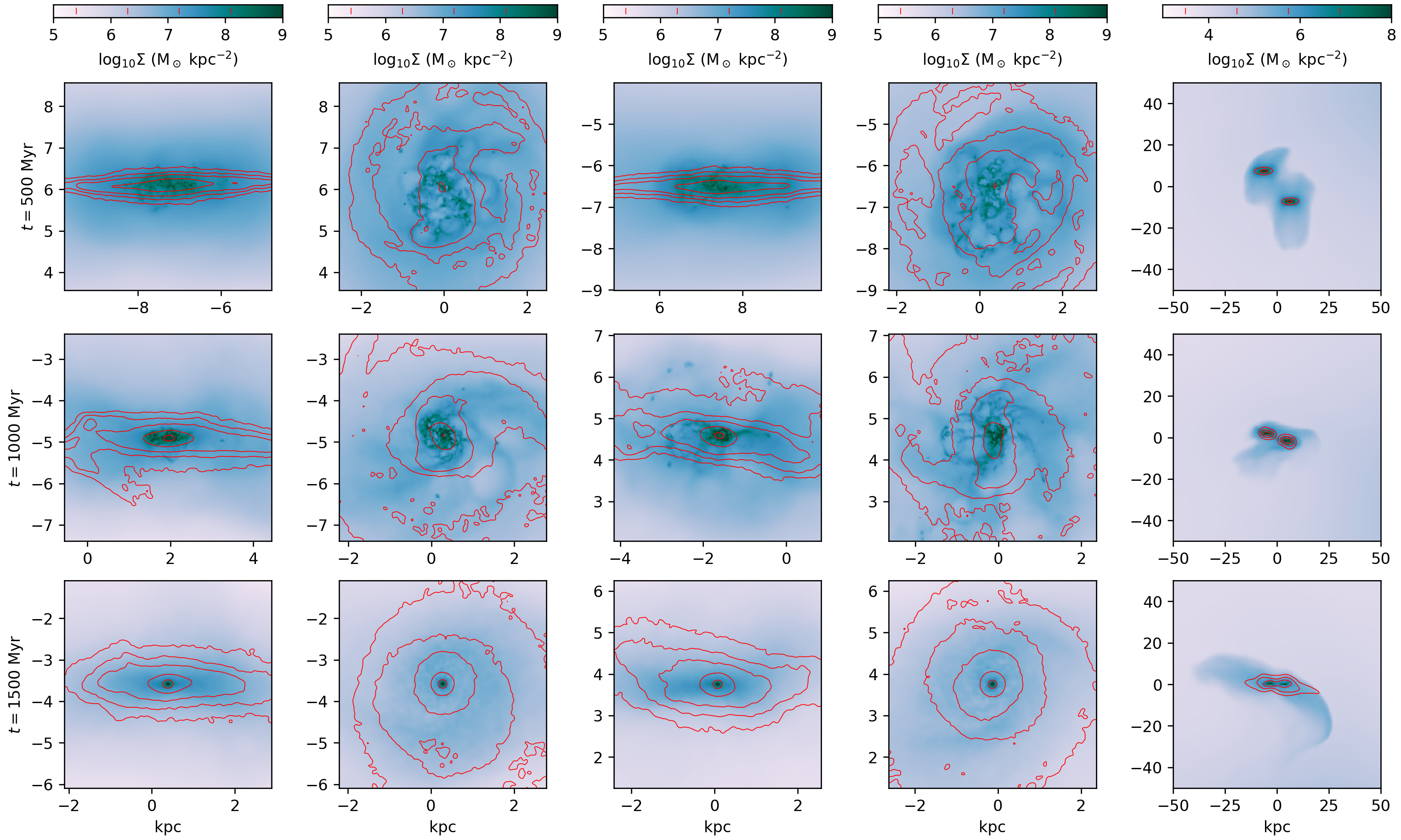}
\end{center}
\caption{\label{G_evolution}
Same as Figure~\ref{D_evolution}, for Run~G}
\end{figure*}

The ``effective yield'' of star formation is plotted in the fourth row of Figure~\ref{timesummary2}. We define this as:

\begin{equation}
\eta_\mathrm{eff}(t)=\frac{Z_g(t)-Z_g(0)}{M_{\rm SF}(t)},
\end{equation}
where $Z_g(t)$ is the total mass of metals (of all tracked species) in the gas phase within a $10$ kpc distance from the center of each galaxy, $M_{\rm SF}(t)$ is the total mass of stars formed since the beginning of the simulation, and $\eta_\mathrm{eff}(t)$ is the effective yield. This value measures the production of metals from star formation, as well as the ejection of metals through feedback. Note that the very large yields at very early times are not representative, but are caused by the initial distribution of ages and masses of stars -- stars present in the initial conditions at $t=0$ produce metal enrichment before any new stars have formed.

After about $500$ Myr, the yield in all simulations reaches a steady state of $\eta_\mathrm{eff}\sim0.02$. Here, star formation is ongoing but outflows are not yet strong enough to carry much metals out of the galaxy. At this point, the rate of increase of metal mass is simply proportional to the star formation rate. The effective yield therefore becomes close to the `true' yield we might expect in a closed box, which observations have shown to be $\sim10^{-2}$ (\citealt{2002ApJ...581.1019G,2004ApJ...613..898T}, and see also the sample yields for GCD+ in Fig.~2 of \citealt{2003MNRAS.340..908K}). 

Rapid star formation then produces a metal-rich outflow at $t\sim1000\,\Myr$, causing the effective yield to drop. This effect is much more significant in Runs~B--E, where mergers trigger strong bursts of star formation and feedback. After this, continual star formation causes the effective yield to slowly recover back towards the steady state. In Runs~A and F, the weak outflows caused by weak star formation causes only a small drop in effective yield at this time, and the effective yield remains closer to the true yield.

Figure~\ref{profiles} shows the disk-plane radial profiles of the abundance ratios \FeH\ and \OFe\ of each dwarf, or their merger remnant, at various epochs. Note that in the runs resulting in mergers (B, C, D, and~E), the system is not fully relaxed by the end of the simulation. These profiles are centered on the minimum of the potential well, which does not exactly correspond with the location of the metallicity peak, resulting in the little ``glitches'' seen in the central parts of the profiles at late time. 

In all runs, star formation is more intense in the center of the dwarfs than in the outer regions, resulting in more metal production in the center, and negative \FeH\ slopes. As star formation continues, metallicities continue to increase in the center. The final peak metallicity in the dwarf centers generally relates to the strength of star formation. Runs~B--D produce a higher peak metallicity than Runs~A and~F. This star formation is trigged by mergers, and the runs without mergers show smaller high-metallicity cores. The peak metallicities in the final snapshot are quite high, which may relate to our feedback method, as we discuss below.

The metallicity slopes at $t=2500$ Myr are quite steep in Runs~A and~F, but are much shallower in the merger remnants of Runs~B--E. Here, the merger has disturbed and mixed the gas, producing a flatter metallicity slope -- although metals are still somewhat centrally concentrated. Although we do include metal diffusion in the simulations, the dominant effect is the direct advection of metal-rich gas through flows driven by feedback and tidal stirring. Some star formation occurs in outer regions, but the main form of enrichment beyond the central few parsecs is from bulk transport of metal from the inner region.

The evolution of the \OFe\ profiles relative to the \FeH\ profiles is primarily determined by the time delay between the onset of Type~II SNe (which produce oxygen and other $\alpha$-elements), and Type~Ia SNe (which produce most of the iron). The values of \OFe\ are initially high because Type~II SNe explode almost immediately, the lifetime of their progenitors being very short on the scale of the simulations. Eventually, Type~Ia SNe explode, resulting in an increase in iron abundance and a corresponding drop in \OFe. Generally, we see a positive slope develop, due to ongoing star formation producing $\alpha$-enriched gas in the dwarf centers. The observed Magellanic Clouds appear to have flatter slopes, and our models may be underestimate mixing from feedback and turbulence (see Section~\ref{section_discussion})

\subsection{The ``Magellanic'' Stream}\label{sec_stream}

\begin{figure*}
\begin{center}
\includegraphics[width=\textwidth]{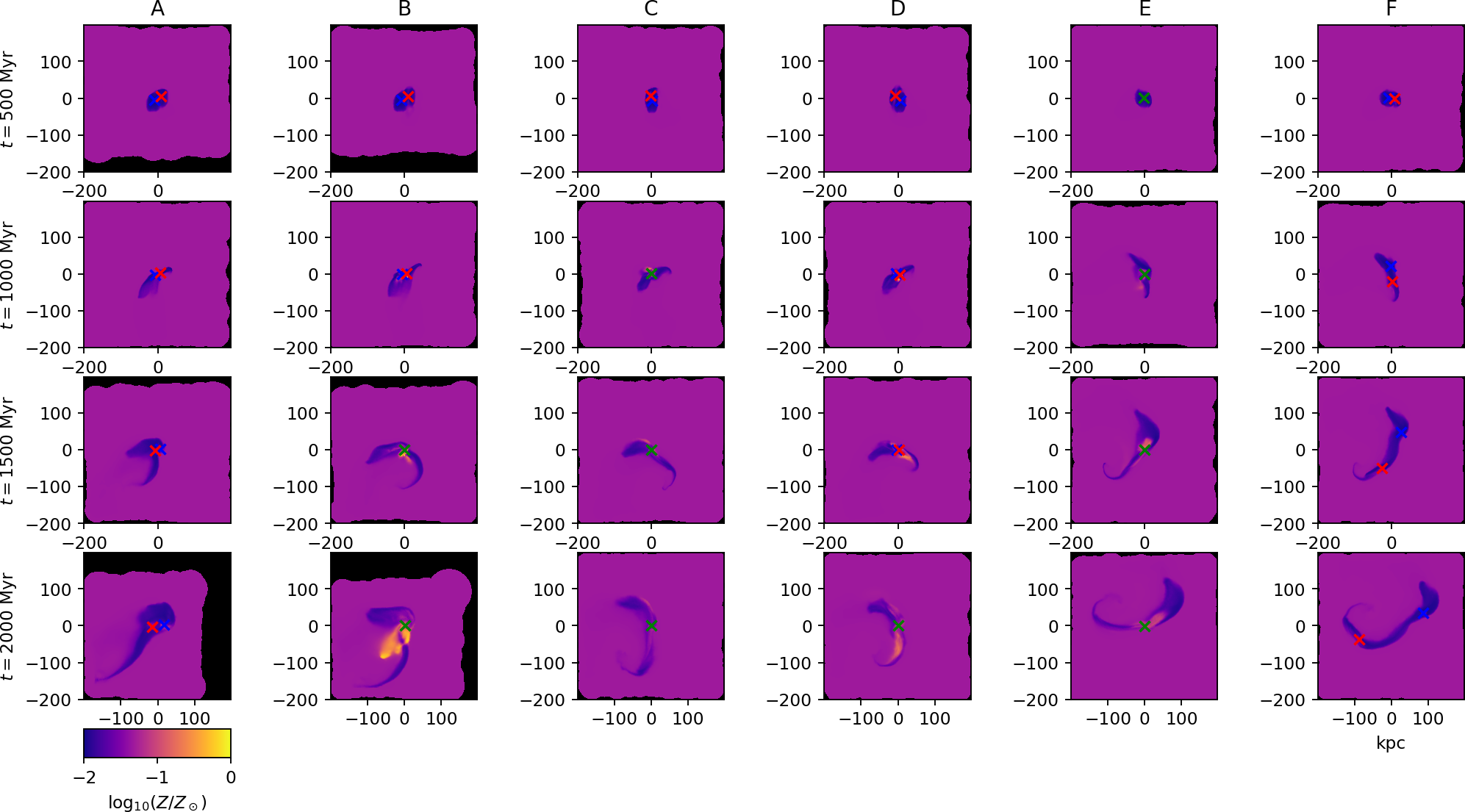}
\end{center}
\caption{\label{metal_maps}
Evolution of the large-scale metal distributions, for Runs~A--F. Blue and red crosses indicate locations of each dwarf, green cross indicates merged dwarf location. The host halo center is at the origin.}
\end{figure*}

\begin{figure*}
\begin{center}
\includegraphics[width=.9\textwidth]{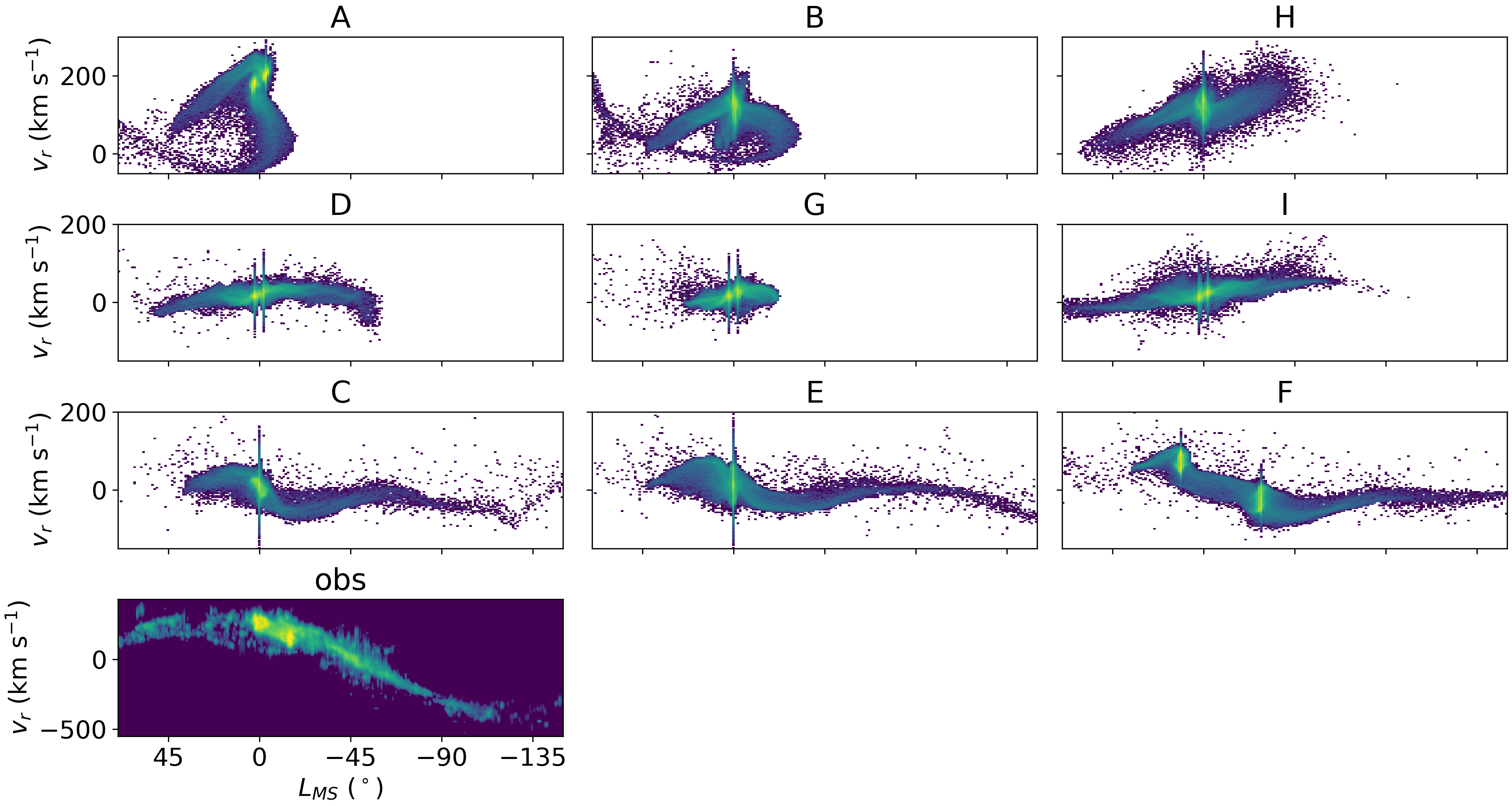}
\end{center}
\caption{\label{dvelstream}
Top three rows: position--velocity diagram for Runs A--I at $t=1500$ Myr. Bottom: observed position--velocity diagram from \citet{2010ApJ...723.1618N}}
\end{figure*}

\subsubsection{Production \& Evolution}

Figures~\ref{column} and \ref{column2} show the evolution of the gas column density for Runs~A--F and Runs~H--K, respectively. We use a two-variable color scheme to illustrate the origin of the gas, with $\Sigma_1$ and $\Sigma_2$ being the column density of gas extracted from the first and second dwarf, respectively. The contribution of gas from the host halo to the stream is very small, and is not plotted here.

The outflows from the two dwarfs produce halos that are initially unmixed, except in the region where the halos directly 
overlap -- the halos quickly become larger than the inter-dwarf distance. As the dwarfs orbit 
around the host galaxy, their gas halos are stretched by tidal forces into streams. In the absence of a host galaxy (Runs J \& K, Figure~\ref{column2}), the outflows produce a largely spherical halo instead. Both arms of the streams tend to contain gas from both dwarfs. Although largely well-mixed, typically one tail of the stream consists of somewhat more gas from one dwarf than the other. The dominant dwarf for each tail is simply that dwarf that is on that side of the system -- as our dwarfs have the same mass, no one dwarf is preferentially stripping the other. This effect is particularly strong where mixing is weak -- in Runs~A and~B, which have elliptical orbits and in Run~F, where the dwarfs escape. Some mixing does occur -- the tails in Runs~A and~B are less mixed at $t=1000\,\rm Myr$ than at $t=2000\,\rm Myr$ -- but the difference is still visible by the end of the simulation, where we can clearly see which arm came from which dwarf.

The difference in the level of mixing between the simulations is due to how rapidly the dwarfs orbit each other. In all runs, the dwarfs are initially separated from each other by the same distance, and with two different relative speeds ($30$ km\,s$^{-1}$ and $50$ km\,s$^{-1}$). However, the dwarf-dwarf orbit can be modified by the tidal influence of the host galaxy, with a strength depending on the orbit of the dwarfs through the halo, and this influence tends to dominate over small variations in the relative velocities of the dwarfs. In Runs~C--E, the clouds orbit each other rapidly, producing a mixed outflow stream (see also Figure~\ref{timesummary}). In Runs~A and~B, the tidal force of the host increases the period of the dwarf-dwarf orbits, and there is less mixing. Similarly, in Run~F, the relative orientation of the dwarf-dwarf orbit allows tidal forces to separate the dwarfs from each other, and mixing is reduced. 

All simulations produce two steams stretching in opposite directions from the dwarfs, a leading one and a trailing one, similar to the LA and TA found in the Magellanic system. However, these labels are somewhat arbitrary and ambiguous. For example, in Run A, a leading and a trailing arm are formed, but the dwarfs overtake the leading arm in their orbit. This causes the leading arm to become a trailing arm. Meanwhile, the trailing arm swings around to become a leading arm. The two arms effectively swap places. To avoid ambiguity, we define the leading and trailing arms according to the velocities and morphology of the system at the snapshot of interest -- i.e. the `leading arm' is the arm currently closest to the dwarfs' direction of motion, regardless of its historical origin.

In Run~F, where the dwarfs end up moving apart at late times, we note the persistence of the bridge between them, that stretches over a distance of $200\,\kpc$. This further demonstrates how tidal forces dominate over ram pressure in these simulations - the bridge is stretched by tides, and is not significantly stripped by ram pressure.

\subsubsection{Effects of Ram Pressure on Evolution}

Ram pressure does have some effect, particularly in Runs~A and~B, where the dwarfs plunge the deepest into the host halo. We should expect ram pressure to push gas in the opposite direction of the dwarfs motion (yellow arrows in Figures~\ref{column}-\ref{column2}). By $t=1500$ and $t=2000$ Myr, the leading arms of the streams do appear to be bent backwards. This confinement of the leading arm is also visible (but less dominant) in some others runs. Runs H \& I, which lack any ram pressure, show similar stream structures to Runs B \& D, although the streams of Runs H \& I are more extended and diffuse. Here, the effect of the host gas halo is not to produce streams through ram pressure, but to constrain the outflows into narrow streams, and to somewhat limit the distance they propagate to. Interestingly, in some runs there are perhaps indications of a two-tailed trailing arm, where one tail is produced by tides, and the other is produced by ram pressure. This is most clear in Run~D at $t=1500$ Myr, where the fork of the trailing arm is directly opposite the dwarfs' direction of motion and appears to be influenced by ram pressure.

\begin{figure*}
\begin{center}
\includegraphics[width=.9\textwidth]{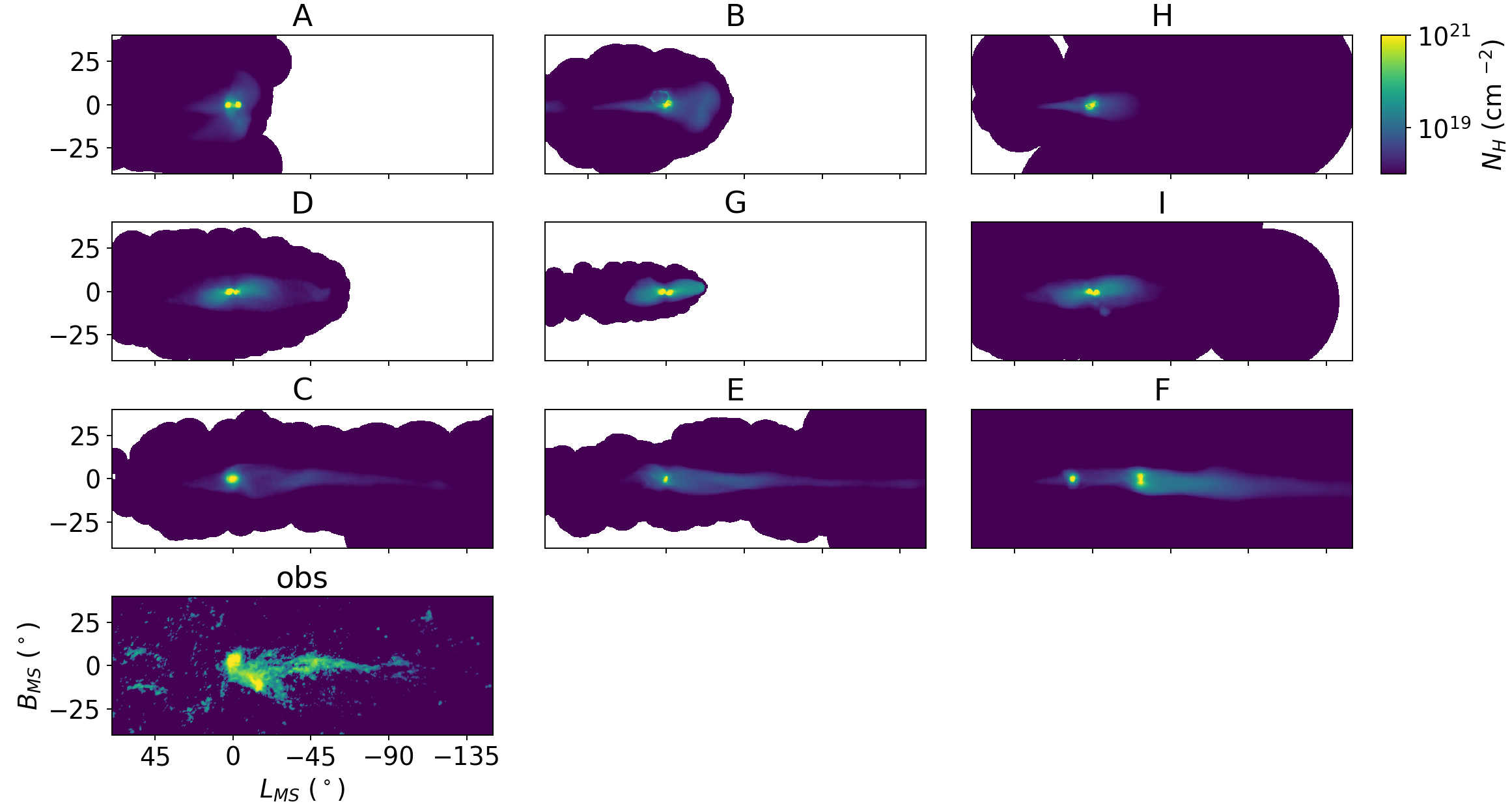}
\end{center}
\caption{\label{earthframe}
Top three rows: Column density maps for Runs A--I at $t=1500$ Myr, as seen from host galaxy centre. Bottom: observed HI column density map from \citet{2010ApJ...723.1618N}}
\end{figure*}

We performed one additional run, Run~G, at a higher host halo gas density, closer to that of the Milky Way as measured by \citet{2018ApJ...862....3B}. This run uses the same orbits as Run~D, the only difference being the mass of the host galaxy (see Table~1).
A visualization of the evolution of the system for this run is shown in Figure~\ref{G_evolution}.
To limit computational cost, we stopped this simulation at $t=1700\,\rm Myr$.

Comparing Figures~\ref{D_evolution} and~\ref{G_evolution}, there is no systematic differences in the internal evolution of the dwarfs. Both show similar disturbances, with only stochastic differences. We also find that star formation rates are very similar.

The key differences are in the large-scale evolution of the streams. Here we again find that pressure {\em confinement} has a stronger effect than ram pressure stripping. A leading and a trailing arm are still both present, and largely point in the same direction as in Run D. Ram pressure stripping has a small effect, in that these arms are somewhat curved `backwards' at the ends. But the more significant effect is that the arms do not extend quite as far beyond the dwarfs. Increasing the gas density by a factor of $10$ has approximately halved the extent of the arms. This is caused by the pressure of the gas in the halo resisting the outflows and tidal forces that stretch out the MS.

We emphasize that pressure confinement constrains \emph{both} the leading and trailing arm. This differs from ram pressure stripping, which enhances the trailing arm but constrains (or prevents) any leading arm.

\subsubsection{Properties}

Figure~\ref{metal_maps} shows the time evolution of the gas metallicity, for Runs~A--F. The upper limit of the scale corresponds to solar abundance. As the two dwarfs are initially identical, we should not expect that metallicity can uniquely trace a stream to its source. Instead, the streams generally have low metallicities. They largely trace their origin to the gas in the low-metallicity outer regions of the dwarfs  (section~\ref{chemo_section}) that has been entrained by outflows. As the streams take time to propagate, and the DGs are continuing to produce metals through star formation, the streams are sampling the lower metallicity of an earlier era in evolution of the DGs. However, a strong burst of star formation can also produce a high-metallicity outflow that adds an enriched component to the stream. The resulting streams are a combination of low-metallicity and high-metallicity gas, that does not exactly match the metallicity of either dwarf galaxy. This highlights the difficulty in attempting to trace the origin of the MS by comparing abundances to those of the MCs. 

The metallicity gradients of the simulated streams are also not uniform, even in sign. These gradients depend on the nature of the outflow. When the outflow is gentle (see Run~D and Run~E at $t=2000$ Myr), the metallicity gradient is negative as enriched gas mixes with the metal-poor medium as it flows. When the outflow is dramatic (see Run~B at $t=2000$ Myr), the metallicity gradient is positive. Here, an enriched front ploughs through the medium, leaving behind a lower density wake (see Figures~\ref{column}-\ref{column2}), that mixes with the metal-poor medium. However, in general, our simulated MS shows lower metallicities than the DGs, consistent with observations \citep{2018ApJ...854..142F}.

We also examine the kinematics of these systems. We calculate the line-of-sight velocity ($v_r$) and position in Magellanic coordinates \citep{2008ApJ...679..432N}, for all gas particles that originated in the dwarfs, at $t=1500\,\rm Myr$. We define our Magellanic coordinates by setting $L_{MS}=0^\circ$ and $B_{MS}=0^\circ$ at center of mass of the dwarf system, where $L_{MS}$ is the azimuthal position along the orbital plane as seen from the halo center, and $B_{MS}$ is defined along the orthogonal great circle. We plot the $L_{MS}-v_r$ distribution in Figure~\ref{dvelstream}. Here we have reordered our runs so that models with similar orbits can be more clearly compared. Na\"ively, we might expect that tides and ram pressure would produce opposite kinematic signatures -- tides would show acceleration away from the dwarfs, while ram pressure would show deceleration towards the host galaxy velocity in the trailing arm but no leading arm at all. In these simulations that combine tides and ram pressure, we often find multiple kinematic components, with the exact structure and gradient depending on the orbital configuration.

Runs A, B, and H have strong radial velocity gradients. Here, the dwarfs have passed periapsis with the host galaxy, and are on nearly radial orbits out of the system. Run A has a stronger velocity gradient than Run B because the small difference in initial velocities means that the dwarfs in Run A have travelled further on their orbit and therefore are on a more radial trajectory. Run H has the same orbit as Run B, but excludes ram pressure, providing a basis to demonstrate the effects of ram pressure. In this case, tides alone (Run H) produce a simple linear velocity gradient along the streams and through the merged dwarf galaxy. Ram pressure (in Run B) has little effect on the trailing arm, but truncates the leading arm, and brings its velocity down to that of the halo, inverting the velocity gradient seen from tides alone.

The remaining runs have nearly circular orbits, and therefore small radial velocities and radial velocity gradients. Runs~D, G, and I, have the same initial orbital configuration, but Run~D has moderate halo gas density, Run~G has strong halo gas density, and Run~I has no halo gas. The difference between Runs~D  and~I is similar to the difference between Runs~B and~H -- ram pressure causes the simple linear velocity gradient in the tides-only run to be flattened in the leading arm, and causes the leading arm to be somewhat truncated. Run G shows truncation of both the leading and trailing arm. Here, pressure confinement is again clearly dominating over ram pressure stripping.

Runs C, E, and F show a clear S-shape in their velocities. The inner gradient is produced by the rotation of the dwarf-dwarf system, and the outer gradient by tides and host gas pressure (as in Runs D, G, and I). In these three runs, the dwarfs have retrograde orbits at their merger or closest encounter, and the gas expelled during the interaction has an inverted velocity gradient. While the dwarfs in Run C initially have a prograde orbit, the influence of the host tides causes the dwarfs to merge in a retrograde sense. This can be seen in the top row of Figure~\ref{timesummary}. The velocity-gradient from the dwarf-dwarf rotation is less visible in Runs D, G, and I, where the opposite process happens. There, the dwarf orbits are initially in a retrograde orbit, but the host tides cause the dwarfs to merge and eject gas in a prograde fashion.

We can compare our kinematics plot with the position--velocity diagram of \citet{2010ApJ...723.1618N}, which we have also plotted in Figure~\ref{dvelstream}. Here the rotation of the Magellanic system is visible, as is a long linear velocity gradient in the trailing arm that flattens at large distances, which we have shown is consistent with a tidal origin for the trailing arm, somewhat truncated by host halo pressure. The leading arm is short and has a flat velocity gradient. From our simulations, we interpret this as a tidally stripped tail, decelerated by halo pressure.

We plot the gas column density in Magellanic coordinates centered on the center-of-mass of the dwarfs, as seen from the center of the host, in Figure~\ref{earthframe}. This is the total surface density of all gas species, but can be compared qualitatively with the HI maps of \citet{2003ApJ...586..170P} and \citet{2010ApJ...723.1618N}. We include the HI map of \citep{2010ApJ...723.1618N} in Figure~\ref{earthframe}, for comparison. In Runs A, B, \& H, the apparent size of the streams are considerably shorted by projection effects, because the DGs' are on nearly radial orbits and produce nearly radial streams. This foreshortening is particularly noticeable in Run A, where the arms are most radial. While we do produce extended streams, the projected size of these stream is particularly parameter dependent, and (as stated above) we do not attempt to fine-tune our models to produce the exact geometry of the Magellanic System.

In general, the projected column density maps show that most runs produce leading arms truncated by halo pressure, and a somewhat more extended trailing arm. Runs H \& I, without ram pressure, show leading and trailing arms of similar extent. This further supports that tidal stripping with pressure confinement can explain the general extent and morphology of the MS.

\section{Discussion}\label{section_discussion}

We have presented results from a series of simulations that produce a Magellanic-like stream with both leading and trailing arms, even with explicit modeling of host halo pressure. This contrasts with many recent simulations of the Magellanic system that have failed to produce a leading arm. These either used higher MW halo masses in their initial conditions \citep{2019MNRAS.488..918T,2019MNRAS.486.5907W} or neglect tidal forces \citep{2018ApJ...863...49B}. Either the true MW halo density is lower than the value used in these simulations, or the leading arm must have an alternate origin \citep{2019MNRAS.488..918T}. We have concentrated on the former option in this work. While a lower density MW halo may not produce the observed excitation of the MS through hydrodynamic shocks, this excitation could have been produced by past AGN activity in the MW \citep{2015ApJ...813...94T}. A survey of L$_*$ galaxies \citep{2017ApJ...837..169P} favors a more massive halo, but observations based on direct measurements of lines of sight within the Milky Way favor a less massive halo \citep{2013ApJ...770..118M,2018ApJ...862....3B}, and X-ray observations suggest that the halo may not even be isotropic \citep{2018ApJ...862...34N}. It may simply be the case that the Milky Way gaseous halo is slightly underdense for an L$_*$ galaxy, at least along part of the trajectory of the MCs.

If we assume a denser MW halo, a possible alternate scenario is that the Magellanic system consists of a sub-group of galaxies \citep[see][]{2015MNRAS.453.3568D} that has been accreted into the MW \citep{2015ApJ...813..110H,2019MNRAS.486.5907W}. In this scenario, one or more of the dwarf galaxies within this structure has led the MCs in their orbit, and has been completely disrupted by tides and ram pressure, to produce a stream of gas leading back to the MCs. However, this may require fine-tuning of parameters: the disrupted dwarf's orbit must align very closely to the MCs, have past abundances close to that of the SMC, and must have had a large reservoir of gas, but have a low enough stellar mass to be disrupted below detectability. A group of tidal dwarf galaxies produced by an encounter with M31 could satisfy these constraints \citep[as noted in][]{2015ApJ...813..110H,2019MNRAS.488..918T,2019MNRAS.486.5907W}, although the LMC would be an unusually massive tidal dwarf galaxy. Suggestions that ram-pressure stripping may also have a weaker effect than expected \citep{2016MNRAS.463.1916F} may also give additional weight to the tidal scenario.

Observations have shown that the leading arm has multiple kinematic components \citep{2012A&A...547A..12V}, something that naturally occurs from the orbits and interactions of the MCs and the MW used in our study (section~\ref{sec_stream}). Our results  
support previous claims that the kinematic structure found in previous tide-only simulations \citep{2018ApJ...857..101P} still largely persists in the presence of a Milky Way gaseous halo. Those simulations, and earlier N-body simulations \citep{2006MNRAS.371..108C,2012ApJ...750...36D} find that tidal stripping is most effective when a more extended MC structure is assumed.

In our simulations, we do not need to make this assumption, since an extended MC structure naturally occurs as a result of feedback-induced outflows triggered by encounters between the MCs. The role of feedback is also emphasized by \citet{2015ApJ...813..110H}. The importance of feedback is further supported by the strong multiphase outflows that have been predicted and observed in the SMC, tied to bursts of star formation $\sim25-60$ Myr ago \citep{2012MNRAS.421.3522H,2018NatAs...2..901M}.

We produce significant metallicity slopes in our dwarf galaxies, with strong central peaks of metallicity. The LMC is observed to have a flatter metallicity slope of order $0.05$ dex/kpc \citep{2009A&A...506.1137C,2016MNRAS.455.1855C}. The SMC may have a similar metallicity slope \citep{2018MNRAS.475.4279C}, although it has been difficult to detect (e.g.~\citealt{2015MNRAS.449.2768D}). Our models also have a higher central metallicity than the SMC and LMC. This is possibly caused by our feedback algorithm, which does not include strong ``preheating'' (through e.g. ionising radiation and radiation pressure) prior to supernova detonation, and may therefore be somewhat underefficient at reducing the central star formation rate \citep{2012MNRAS.421.3488H}. 

\section{Conclusions}\label{section_conclusion}

We have extended our earlier work on the evolution of dwarf galaxies orbiting inside the gaseous halo of a massive host galaxy to the case of Magellanic-like systems consisting of two interacting dwarf galaxies. These simulations include the pressure and tidal forces of the host galaxy. We considered the case of equal-mass dwarf galaxies with masses intermediate between the LMC and SMC, and performed a series of eleven chemodynamical simulations. We vary the orbital parameters of the dwarf galaxies between the simulations, and vary the MW halo density in one simulation. Our results are the following:

\begin{itemize}

\item In all simulations, the system develops both a trailing and a leading gaseous stream, akin to the Magellanic Stream observed in the Milky Way-Magellanic Cloud system. These often show multiple kinematic components with various gradients, depending on the orbital configuration. The streams are produced from gas ejected by bursts of star formation and interaction between the dwarfs, that are then stretched into streams by host galaxy tides, and somewhat truncated by halo pressure. Our model qualitatively reproduces the key features of the observed kinematics and morphology of the Magellanic Stream.

\item The gas of the stream has a mix of high and low metallicities, due to enriched outflows entraining gas from the low-metallicity outer part of the dwarfs. The streams further differ in chemical abundance from the dwarfs because the dwarfs have continued to increase their metallicity through star formation, and because the streams trace their origin from multiple events in different times and and places in the Magellanic Clouds. This demonstrates that direct comparison between Magellanic Stream abundances and those of the Magellanic Clouds -- even in the past -- should be performed with caution.

\item In all simulations, the dwarf galaxies experience one
or several starbursts. The timing of these starburst correspond to the epochs when the galaxies merge or reach their lowest separation.

\item A gaseous bridge forms between the dwarfs. The dwarfs also show the classic signs of disturbance, such as spiral features and asymmetric arms.

\item Pressure confinement dominates over ram-pressure stripping. Increasing the host galaxy gas halo density constrains the stream closer to the dwarfs, rather than stripping additional material.

\item We find that star formation and internal processes largely dominate the evolution of the dwarfs themselves. Interactions between the dwarfs contribute as a secondary effect by distorting their morphologies, and by driving bursts of star formation and feedback. The host galaxy contributes only weakly by perturbing the dwarfs orbits around each other, and only has a significant effect on extended material produced by outflows.
    
\end{itemize}

We conclude by pointing out that strong feedback and a moderate host halo gas density can reproduce the qualitative characteristics of the Magellanic Stream, including the leading arm, solving a major problem with models where feedback is too weak or the host halo gas density is too high. Feedback-driven outflows extract gas from the Magellanic Clouds. Tides stretching these outflows into a stream. Host halo pressure somewhat truncates these streams, but does not strip gas directly from the Magellanic Clouds, nor prevents the formation of a leading arm.

\section*{Acknowledgments}DJW is supported by European Research Council Starting Grant ERC-StG-677117 DUST-IN-THE-WIND. HM is supported by the Natural Sciences and Engineering Research Council of Canada. We acknowledge significant usage of the following (pip-installable) Python libraries: pykdgrav, pynbody, matplotlib, scipy, numpy, pandas. We thank David Nidever for providing data from \citet{2010ApJ...723.1618N}.

\bibliographystyle{apj}
\bibliography{double_dwarf}

\end{document}